\documentclass[12pt,a4paper]{article}
\usepackage{epsfig}
\newcommand{\beqar}{\begin{eqnarray}}
\newcommand{\eeqar}[1]{\label{#1} \end{eqnarray}}

\title{ Nuclear Mean Fields through Selfconsistent Semiclassical 
Calculations\footnote{This work is supported in part by the French-Algerian 
CNRS/DEF scientific research agreement under Contract No. PNC 10090}}

\author{
J.~Bartel  \\
{\it Institut de Recherches Subatomique} and {\it Universit\'e Louis Pasteur} \\
{\it Strasbourg, France}                     \\
                                                                     \\[ 3.0ex]
K.~Bencheikh  \\
{\it Departement de Physique, Facult\'e des Sciences}, \\
{\it Universit\'e de Setif, Setif 19000, Algeria}
                                                                     \\[ 3.0ex]
}
\begin{document}
\noindent
\maketitle
\vspace{1cm}
\begin{abstract}
Semiclassical expansions derived in the framework of the Extended Thomas-Fermi 
approach for the kinetic energy density $\tau(\vec{r})$ and the spin-orbit 
density $\vec{J}(\vec{r})$ as functions of the local density $\rho(\vec{r})$ 
are used to determine the central nuclear potentials $V_n(\vec{r})$ and 
$V_p(\vec{r})$ of the neutron and proton distribution for effective 
interactions of the Skyrme type. 
We demonstrate that the convergence of the resulting semiclassical expansions 
for these potentials is fast and that they reproduce quite accurately the 
corresponding Hartree-Fock average fields.
\end{abstract} 

\newpage
\section{Introduction}
\vspace{-0.2cm}

Mean-field calculations have been extremely successfull over the last 3 decades 
to describe the structure of stable as well as radioactive nuclei and this over 
a very wide range of nuclear masses. Especially effective nucleon-nucleon 
interactions of the type of Skyrme \cite{Sk59,VB72} and Gogny forces 
\cite{DG80} have been particularly efficient in this context. Such 
phenomenological effective interactions can be understood as mathematically 
simple parametrisations of a density-dependent effective G-matrix (see 
\cite{Ne82} and \cite{QF78} for a review on such effective forces). 

Together with the exact treatment of the mean-field problem in the 
Hartree-Fock (HF) approach, semiclassical approximations thereof have proven 
very appropriate. Especially the approach known as the Extended Thomas-Fermi 
(ETF) method has been shown \cite{BGH85} to describe very accurately average 
nuclear properties in the sense of the Bethe-Weizs\"acker mass formula 
\cite{We35,Be36}. In their selfconsistent version the ETF calculations 
determine the structure of a given nucleus by minimizing the total energy with 
respect to a variation of the neutron and proton densities. 
Such calculations require, however, only {\bf integrated} quantities as e.g.
the total nuclear energy. 

The aim of the present paper is, on the contrary, to investigate {\bf local} 
quantities such as nuclear mean-field potentials, effective mass and spin-orbit 
form factors which are at the basis of the description of the nuclear structure 
and which can be obtained as function of the selfconsistent semiclassical 
densities. The convergence of these local (non integrated) 
semiclassical quantities and their comparison to the corresponding HF 
distributions has, to our knowledge, never been extensively investigated 
as this will be done here.

This paper is organized as follows. After specifying in section 2 the ETF 
expressions up to order $\hbar^4$ for $\tau[\rho]$ and $\vec J[\rho]$ 
using the general but rather 
cumbersome form of these expressions given in \cite{GV79,GV80}, we show in 
section 3 that the semiclassical expansions which define these quantities 
converge rapidly for reasonable forms of the nuclear density $\rho(\vec r)$. 
Once these expressions and their convergence established, we compare in 
section 4 the average mean fields obtained using these semiclassical densities 
to the central potentials resulting from a HF calculation. 
We finally conclude giving an outlook on how these calculations can be 
generalized to excited and rotating nuclei.
%
%
%
\section{Form factors for Skyrme Interactions}
\vspace{-0.2cm}

For effective nucleon-nucleon interactions of the Skyrme type the total energy 
of a nucleus is a functional of the local densities $\rho_q(\vec{r})$, 
the kinetic energy densities $\tau_q(\vec{r})$ and the so called spin-orbit 
densities $\vec{J}_q(\vec{r})$ \cite{VB72}
\begin{equation}
    E = \int {\cal E} ( \rho_q(\vec{r}), \tau_q(\vec{r}), \vec{J}_q(\vec{r})) 
        \; d^3r \; ,
\label{c2.1}\end{equation}
where the subscript $q \!\!=\!\! \{n,p \}$ denotes the nucleon charge state.
In the case of broken time-reversal symmetry the energy density depends, in 
addition, on other local quantities \cite{EBG75,BBQ94}, such as the current 
density $\vec{j}(\vec{r})$ and the spin-vector density $\vec{\rho}(\vec{r})$. 
In what follows we will restrict ourselves to time-reversal symmetric systems 
leaving the case of broken time-reversal symmetry, particularly 
encountered in the case of rotating nuclei, to a subsequent publication. 

The total energy determined in this way is exact within the HF formalism. A 
semiclassical approximation thereof is obtained when instead of using the 
exact quantum-mechanical densities $\rho_q(\vec{r})$, $\tau_q(\vec{r})$ 
$\vec{J}_q(\vec{r})$, etc. a semiclassical approximation for these quantities 
is used. The semiclassical densities $\tau_q(\vec{r})$ and $\vec{J}_q(\vec{r})$ 
are obtained in the so called 
{\it Extended Thomas-Fermi model} \cite{BGH85} as functions of the local 
density $\rho(\vec{r})$ and of its derivatives. The best known of these 
functionals is the Thomas-Fermi approximation for the kinetic energy density 
in the form 
\begin{equation}
    \tau_q^{(TF)}[\rho_q] = \frac{3}{5} \, (3 \pi^{2})^{2/3} \, \rho_q^{5/3} 
                                     \;\; , \;\;\;  q \!\!=\!\! \{n,p \} \; .
\label{c2.2}\end{equation}
Once these functional expression are given, the total energy of the nuclear 
system is then uniquely determined by the knowledge of the local densities 
of protons and neutrons. 
That such a functional dependence of the total energy on the local densities 
$\rho_q(\vec{r})$ should exist is guaranteed by the famous theorem by 
Hohenberg and Kohn \cite{HK64}. In the most general quantum mechanical case 
this functional is, however, perfectly unknown. The great advantage of the 
semiclassical approach used here, consists in the fact that, in connection with 
effective interactions of the Skyrme type such an energy functional ${\cal E}$
can be derived explicitly. In addition, it is to be noted that the 
semiclassical functionals obtained in the ETF formalism such as $\tau[\rho]$ 
are completely general and valid for any local interaction and any nucleus, 
and can therefore be determined once and forever. 

Once the functional of the total energy is known, one is able, due to the 
Hohenberg-Kohn theorem, to perform density variational calculations, where the 
local densities $\rho_q(\vec{r})$ are the variational quantities. One should, 
however, keep in mind that, as the ETF functionals are of semiclassical nature, 
the density functions $\rho_q(\vec{r})$ obtained as a result of the variational 
procedure can only be semiclassical in nature, i.e. of the liquid-drop type. 
Taking into account that in such a process the particle numbers $N$ and $Z$ 
should be conserved, one can formulate the variational principle in the form 
\begin{equation}
   \frac{\delta}{\delta \rho_q} \int \left\{ ^{}
    {\cal E}[\rho_n(\vec{r}), \rho_p(\vec{r})] - \lambda_n \rho_n(\vec{r}) 
                                     - \lambda_p \rho_p(\vec{r}) \right\} d^3r 
\label{c2.3}\end{equation}
with the Lagrange multipliers $\lambda_n$ and $\lambda_p$ to ensure the 
conservation of neutron and proton number. 

This density variational problem has been solved 
in two different ways in the past: either by resolving the 
Euler-Lagrange equation \cite{BBD85,CPV90} resulting from eq.\ (\ref{c2.3}) 
or by carrying out the variational calculation in an restricted subspace of 
functions adapted to the problem, i.e. being of semiclassical nature, 
free of shell oscillations in the nuclear interior and falling off 
exponentially in the nuclear surface. It has been shown \cite{BGH85,BBD85} 
that modified Fermi functions which for spherical symmetry take the form 
\begin{equation}
   \rho_q(\vec{r})] = \rho_{0_q} \, 
                   \left[ 1 + exp(\frac{r - R_{0_q}}{a_q}) \right]^{-\gamma_q}
                                    \;\; , \;\;\;  q \!\!=\!\! \{n,p \} 
\label{c2.3a}\end{equation}
are particularly well suited in this context and that the semiclassical 
energies obtained are, indeed, very close to those resulting from the exact 
variation \cite{BBD85,CPV90}.
\begin{center}
\vspace{-0.2cm}
\hspace*{0.2cm}\epsfig{file=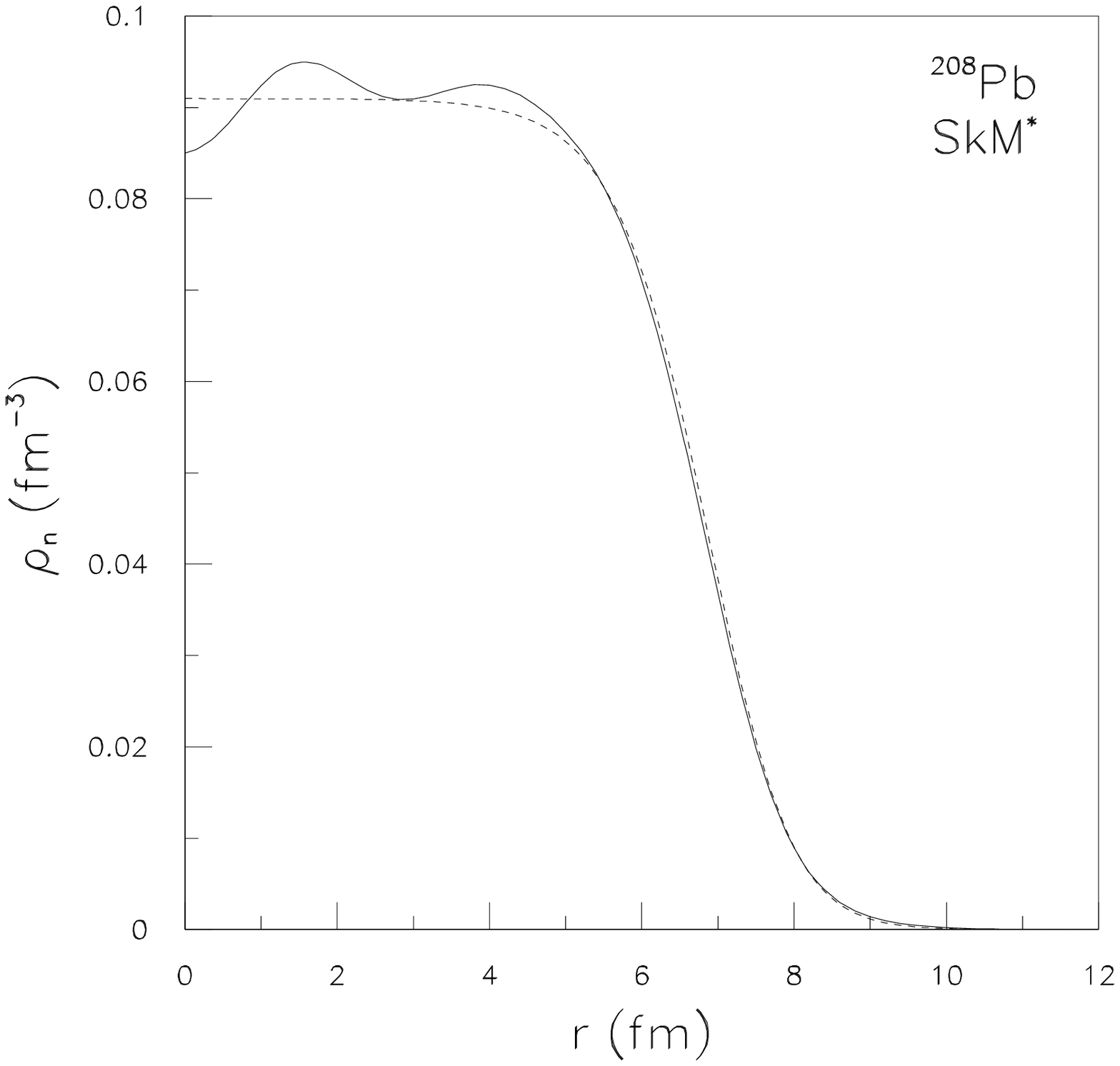,height=8.0cm,width=14.0cm}  \\[ -8.0ex]
\hspace*{0.2cm}\epsfig{file=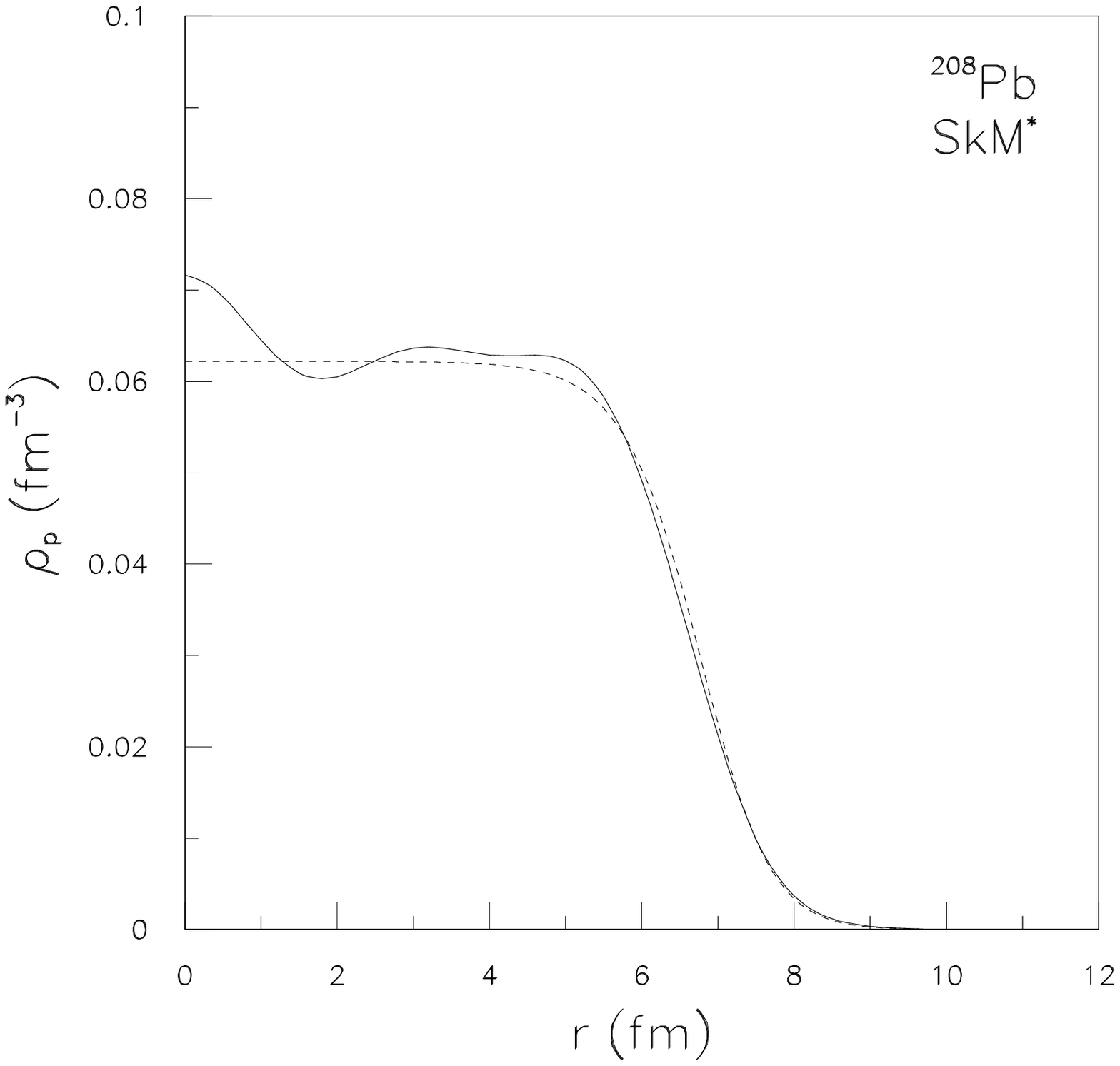,height=8.0cm,width=14.0cm}
\vspace*{-0.8cm}
\begin{small}
\begin{itemize}
\item[]
Fig. 1. Comparison of selfconsistent neutron and proton HF (solid line) and
ETF densities (dashed line) for $^{208}$Pb calculated with the SkM$^*$ Skyrme
force.
\end{itemize}
\end{small}
\end{center}
\vspace{0.2cm}

As an example of the quality of the semiclassical density obtained by such a 
restricted variation in the subspace of modified Fermi functions, eq. 
(\ref{c2.3a}), we show in Fig. 1 a comparison of the neutron and proton 
densities obtained in this way within the ETF approach with the corresponding 
Hartree-Fock (HF) densities for the nucleus $^{208}$Pb calculated with the 
Skyrme interaction SkM$^*$ \cite{BQB82}. 
It should be emphasized here that a similarly good agreement as the one shown 
in Fig. 1 is obtained for other nuclei or using other effective interactions 
such as the Skyrme forces SIII \cite{BFN75} and SLy4 \cite{CBH97}.

The energy density ${\cal E}$ appearing in eqs. (\ref{c2.1}) and (\ref{c2.3})
can be written for a Skyrme interaction as defined in ref.\ \cite{GQ80} 
in the compact form \cite{BFH87}
\begin{eqnarray}
 &&   {\cal E}(\vec{r}) = \frac{\hbar^2}{2m} \tau 
                + B_1 \rho^2 + B_2 (\rho_n^2 + \rho_p^2) 
                + B_3 \rho \, \tau + B_4 (\rho_n \, \tau_n + \rho_p \, \tau_p)
                                                                    \nonumber\\
 &&             - B_5 (\vec{\nabla} \rho)^2 
        - B_6  \left[ (\vec{\nabla} \rho_n)^2 + (\vec{\nabla} \rho_p)^2 \right]
                + \rho^{\alpha} [ B_7 \rho^2 + B_8 (\rho_n^2 + \rho_p^2) ]
                                                                    \nonumber\\
 &&             - B_9 \left[ \vec{J} \cdot \vec{\nabla} \rho 
                           + \vec{J_n} \cdot \vec{\nabla} \rho_n 
                           + \vec{J_p} \cdot \vec{\nabla} \rho_p \right] 
                + {\cal E}_{Coul}(\vec{r}) \; ,
\label{c2.6}\end{eqnarray}
given in terms of the coefficients $B_1 \!-\! B_{9}$ (see table 1) instead of 
the usual Skyrme force parameters $t_0,t_1,t_2,t_3,x_0,x_1,x_2,x_3,W_0$~:
                                                                    \\[ -3.0ex]
\begin{center}
\begin{tabular}[h]{|lll|lll||ll|ll|}
\hline 
 &&&&&&&&&                                                          \\[ -1.0ex]
  &  $B_1$  &  & 
  &  $\frac{1}{2} t_0 (1 + \frac{x_0}{2})$  &  & $\;$  
  $B_6$ &  &  &  $ \frac{1}{16} \left[3 t_1 (x_1 \!+\! \frac{1}{2}) 
                                  + t_2 (x_2 \!+\! \frac{1}{2}) \right] \; $
                                                                    \\[  0.8ex]
 & $B_2$ &   &   & $-\frac{t_0}{2} (\frac{1}{2} + x_0)$  &   & $\;$
   $B_7$ &   &   & $ \frac{1}{12} t_3 (1 \!+\! \frac{x_3}{2})$ 
                                                                    \\[  0.8ex]
 & $B_3$ &   &   & $\frac{1}{4} \left[t_1 (1 + \frac{x_1}{2}) 
                                 + t_2 (1 + \frac{x_2}{2}) \right] $  &  & $\;$
   $B_8$ &   &   & $-\frac{1}{12} t_3 (\frac{1}{2} + x_3)$ 
                                                                    \\[  0.8ex]
 & $B_4$ &   &   & $-\frac{1}{4} \left[t_1 (x_1 \!+\! \frac{1}{2}) 
                             - t_2 (x_2 \!+\! \frac{1}{2}) \right] $  &  & $\;$
   $B_9$ &   &   & $-\frac{1}{2} W_0$ 
                                                                    \\[  0.8ex]
 & $B_5$ &   &   & $-\frac{1}{16} \left[3 t_1 (1 + \frac{x_1}{2}) 
                                     - t_2 (1 + \frac{x_2}{2}) \right] $
                                                          &  &  $\;$  &&&
                                                                    \\[  2.5ex]
\hline 
\end{tabular}
\end{center}
\begin{small}
\begin{itemize}
\item[]
Tab. 1. Correspondence between the coefficients $B_i$ used in the text and 
the usual Skyrme force parameters.
                                                                    \\[ -1.0ex]
\end{itemize}
\end{small}
In eq. (\ref{c2.6}) non-indexed quantities like $\rho$ correspond to the sum 
of neutron and proton densities as $\rho \!=\! \rho_n + \rho_p$ and 
${\cal E}_{Coul}$ is the Coulomb energy density which can be written as the 
sum of the direct and an exchange contribution, the latter being taken into 
account in the Slater approximation \cite{Sl51,Go52}
\begin{equation}
   {\cal E}_{Coul}(\vec{r}) = \frac{e^2}{2} \rho_p(\vec{r}) \int d^3r' \, 
        \frac{\rho_p(\vec{r \,'})}{\vert\, \vec{r} - \vec{r \,'} \,\vert}
      - \frac{3}{4} e^2 (\frac{3}{\pi})^{1/3} \,  \rho_p^{4/3}(\vec{r}) \; .
\label{c2.5}\end{equation}

The HF equation is obtained through the variational principle which states 
that the total energy of eq.\ (\ref{c2.1}) should be stationary with respect 
to any variation of the single-particle wave functions $\varphi^{(q)}_j$~:
\begin{equation}
   \hat{\cal H}_q \, \varphi^{(q)}_j = \left( 
      -\vec{\nabla} \frac{\hbar^2}{2 m_q^*(\vec r)} \vec{\nabla} + V_q(\vec r) 
      - i \vec{W}_q(\vec r) \cdot (\vec{\nabla} \times \vec{\sigma}) \right) 
        \varphi^{(q)}_j = \varepsilon^{(q)}_j \, \varphi^{(q)}_j     \; .
\label{c2.7}\end{equation}
Here appear different form factors such as the central one-body 
potential $V_q(\vec{r})$, the effective mass $m_q^*(\vec{r})$ and the 
spin-orbit potential $\vec{W}_q(\vec{r})$ which are all defined as functional 
derivatives of the total energy density. One obtains from eq.\ (\ref{c2.6})~:
                                                                    \\[ -1.5ex]
\begin{eqnarray}
&&
  \!\!\!\!\!\!\!\!\!\!\!\!\!\!\!
   V_q(\vec{r}) = \frac{\delta {\cal E} (\vec{r})}{\delta \rho_q(\vec{r})} = 
                                                                    \nonumber\\
&&
   = 2 ( B_1 \rho + B_2 \rho_q ) + B_3 \tau + B_4 \tau_q 
                  + 2 (B_5 \Delta \rho + B_6 \Delta \rho_q ) 
                  + (2 + \alpha) B_7 \rho^{\alpha + 1} 
                                                                    \nonumber\\
&&  \;\;\;\;\;
                  + B_8 \left[ \alpha \rho^{\alpha - 1} \sum_q \rho_q^2 
                               + 2 \rho^{\alpha} \rho_q \right]
             + B_9 ( \mathrm{div} \vec{J} + \mathrm{div} \vec{J}_q )
                + V_{Coul}(\vec{r}) \, \delta_{pq} \; ,
\label{c2.8}\end{eqnarray}
                                                                    \\[ -3.0ex]
\begin{eqnarray}
&&\hskip-0.50truecm
   f_q(\vec{r}) = \frac{m}{m_q^*(\vec{r})} 
                = \frac{2m}{\hbar^2} \, 
                  \frac{\delta {\cal E} (\vec{r})}{\delta \tau_q(\vec{r})} 
                = 1 + \frac{2m}{\hbar^2} \, 
                        \left[ B_3 \rho(\vec{r}) + B_4 \rho_q(\vec{r}) \right]
\label{c2.9}\end{eqnarray}
and 
\begin{equation}
  \vec{W}_q(\vec{r}) 
                 = \frac{\delta {\cal E} (\vec{r})}{\delta \vec{J}_q(\vec{r})} 
                 =-B_9 \; \vec{\nabla}(\rho + \rho_q)  \; .
\label{c2.10}\end{equation}
The Coulomb potential in eq.\ (\ref{c2.8}) is easily obtained as
\begin{equation} 
  V_{Coul}(\vec r) = e \int \frac{\rho_p(\vec{r \,'})}
                           {\vert\, \vec{r} - \vec{r \,'} \,\vert} \, d^3r \,'
                       - (\frac{e}{\pi})^{1/3} \,  \rho_p^{1/3}(\vec{r}) \; .
\label{c2.11}\end{equation} 
It is noteworthy in this connection that for such an 
effective mass (\ref{c2.9}) and the spin-orbit potential (\ref{c2.10}) 
the energy density (\ref{c2.6}) takes the simple form 
\begin{eqnarray}
 &&   {\cal E}(\vec{r}) = \frac{\hbar^2}{2m} \sum_q f_q \, \tau_q 
                + B_1 \rho^2 + B_2 (\rho_n^2 + \rho_p^2) 
                - B_5 (\vec{\nabla} \rho)^2 
        - B_6  \left[ (\vec{\nabla} \rho_n)^2 + (\vec{\nabla} \rho_p)^2 \right]
                                                                   \nonumber \\
 &&  \;\;\;\;\;\;\;\;\;\;\;\;\;\;\;\;\;\;\;\;\;\;\;\;\;\;\;\;\;\;\;\;  
                + \rho^{\alpha} [ B_7 \rho^2 + B_8 (\rho_n^2 + \rho_p^2) ]
                + \sum_q \vec{J}_q \cdot \vec{W}_q 
                + {\cal E}_{Coul}(\vec{r}) 
\label{c2.12}\end{eqnarray}
which simplifies somewhat the calculation.
                                                                    \\[ -1.0ex]

All the expressions derived so far (eqs. (\ref{c2.5}) -- (\ref{c2.12})) are 
exact and when used with densities constructed from the single-particle wave 
functions $\varphi^{(q)}_j(\vec r)$, solutions of the HF equation (\ref{c2.7}) 
these quantities contain all the quantum effects of the system. 
If one is interested in the semiclassical approximation of these form factors 
one can immediately conclude, from the analytical form of eqs. (\ref{c2.9}) 
and (\ref{c2.10}) and the smooth behavior of the semiclassical densities, as 
demonstrated in Fig. 1, on the smooth behavior of the semiclassical 
effective mass and spin-orbit form factors. As the nuclear quantal density is 
well reproduced on the average by the ETF densities it appears evident 
that the same is going to be the case for the effective mass and the spin-orbit 
potential, when semiclassical, i.e. liquid-drop type densities are used in 
eqs. (\ref{c2.9}) and (\ref{c2.10}). 

Things are, however, less evident for the central nuclear potentials. Not only 
is the functional expression, eq.\ (\ref{c2.8}), much more complicated then 
those for the effective mass and spin-orbit potential, but the central 
potential is also the only of the three functional derivatives that depends 
not only on the local densities $\rho_q(\vec{r})$ and their derivatives but 
also on the kinetic energy density $\tau_q(\vec{r})$ and the spin-orbit density 
$\vec{J}_q(\vec{r})$ which are the quantities for which the Extended 
Thomas-Fermi approach has written down functional expressions. 
We therefore choose to study the convergence of the semiclassical series 
corresponding to these functional expressions of $\tau_q[\rho_q]$ and 
$\vec{J}_q[\rho_q]$ before investigating the quality of the agreement between 
the HF central potential and the one obtained when using these semiclassical 
ETF functionals. 
\vspace{0.1cm}
\section{Convergence of ETF functionals}
%
\vspace{0.1cm}
The semiclassical expansions of kinetic energy density $\tau_q$ and spin-orbit 
density $\vec{J}_q$ as functions of the local density $\rho_q$ are functional 
expressions with $\hbar$ as order parameter. These expressions can be obtained 
for instance through the semiclassical $\hbar$ expansions developed by Wigner 
\cite{Wi32} and Kirkwood \cite{Ki33} or through the semiclassical method of 
Baraff and Borowitz \cite{BB61}. In either of the two approaches one obtains 
functional expressions like 
\begin{equation}
      \tau_q^{(ETF)}[\rho_q] = \tau_q^{(TF)}[\rho_q] + \tau_q^{(2)}[\rho_q] 
                             + \tau_q^{(4)}[\rho_q]    
\label{c2.13}\end{equation}
written here for the kinetic energy density $\tau_q(\vec{r})$ where 
$\tau_q^{(TF)}[\rho_q]$ is the well known Thomas-Fermi expression already 
given in eq.~(\ref{c2.2}), $\tau_q^{(2)}[\rho_q]$ the semiclassical 
correction of second order and $\tau_q^{(4)}[\rho_q]$ is of fourth order in 
$\hbar$. The ETF expressions such as $\tau_q^{(ETF)}[\rho_q]$, eq. 
(\ref{c2.13}), up to order $\hbar^4$ are to be understood as the converging 
part of an asymptotic series. 

The second order term $\tau_q^{(2)}[\rho_q]$ has already been derived in 
ref. \cite{BJC76} for a Hamiltonian, eq.\ (\ref{c2.7}),  with an effective mass 
$m_q^* = m/f_q$ and a spin-orbit potential $\vec W_q$ 
\begin{eqnarray}
&& \tau_q^{(2)}[\rho_q]\!=\!\frac{1}{36} \frac{(\vec{\nabla} \rho_q)^2}{\rho_q} 
        + \frac{1}{3} \Delta \rho_q 
        + \frac{1}{6} \frac{\vec{\nabla} \rho_q \cdot \vec{\nabla} f_q}{f_q} 
        + \frac{1}{6} \rho_q \frac{\Delta f_q}{f_q} 
        - \frac{1}{12} \rho_q \left(\! \frac{\vec{\nabla} f_q}{f_q} \!\right)^2 
                                                                    \nonumber\\
&&  \;\;\;\;\;\;\;\;\;\;\;\;\;\;\;\;\;\;\;\;\;\;\;\;\;\;\;\;\;\;\;\;\;\;\;
        + \frac{1}{2} \left( \frac{2m}{\hbar^2} \right)^2 \rho_q 
                      \left(\! \frac{\vec{W}_q}{f_q} \!\right)^2  
\label{c2.14}\end{eqnarray}
the first term of which is known as the Weizs\"acker correction \cite{We35}. 
It was sometimes used in the past with an adjustable parameter (instead of 
$\frac{1}{36}$) in order to mock up the absence of the other second order and 
all of the fourth order terms. It has, however, been shown (see e.g. ref. 
\cite{BGH85}) that such a procedure is unable to correctly describe both the 
slope of the surface of the nuclear mass or charge density and, at the same 
time, the height of nuclear fission barriers in the actinide region. From 
this analysis we conclude that the inclusion of fourth order terms in the 
semiclassical functionals is, in fact, without credible alternatives. 
                                                                    \\[ -1.2ex]

In the following we are going to exploit the expressions for the 4$^{\rm th}$ 
order functionals $\tau_q$ and $\vec{J}_q$ developed by Grammaticos and Voros 
\cite{GV79,GV80}. These authors have taken the convention ``that any 
free-standing gradient operator acts only on the {\it rightmost} term'' (see 
their remark after eq.\ (III.13) of ref. \cite{GV79}). In our present work we 
prefer to rewrite these terms in a more conventional way and have any free 
standing gradient operator act, as usual, on {\it all} the terms that appear 
on its right hand side. We therefore write (subscripts GV refer to the 
Grammaticos/Voros convention)
\[  \left[ (\vec{\nabla} f_q \!\cdot\! \vec{\nabla})^2 f_q \right]_{GV} 
      = \frac{1}{2} 
        (\vec{\nabla} f_q \!\cdot\! \vec{\nabla}) (\vec{\nabla} f_q)^2  \; , \]
\[  \left[ (\vec{\nabla} \rho_q \!\cdot\! \vec{\nabla})^2 f_q \right]_{GV} 
        = \vec{\nabla} \rho_q \!\cdot\! \vec{\nabla} 
           (\vec{\nabla} f_q \!\cdot\! \vec{\nabla} \rho_q) - \frac{1}{2} 
        \vec{\nabla} f_q \!\cdot\! \vec{\nabla} (\vec{\nabla} \rho_q)^2 \; , \]
and
\[  \left[ (\vec{\nabla} f_q \!\cdot\! \vec{\nabla}) 
    (\vec{\nabla} \rho_q \!\cdot\! \vec{\nabla}) f_q \right]_{GV} = \frac{1}{2} 
        \vec{\nabla} \rho_q \!\cdot\! \vec{\nabla} (\vec{\nabla} f_q)^2 \; , \]
plus terms that are obtained from these ones by interchanging the role of $f_q$ 
and $\rho_q$. One has also to keep in mind that Grammaticos and Voros use a 
slightly different definition of the effective-mass and spin-orbit form factors 
than the ones given in eqs. (\ref{c2.9}) and (\ref{c2.10}) above~:
$$ 
   f_{{}_{GV}} = \frac{1}{m} f
$$
and 
$$
   \vec{S}_{{}_{GV}} = \frac{1}{\hbar^2} \, \vec{W} \; .
$$

Using these expressions we obtain the following form for the 
4$^{\rm th}$ order kinetic energy density, where contributions from the 
spin-orbit interaction have been, temporarily left out. 
\begin{eqnarray*}
&&  \!\!\!\!\!\!\!\!\!
  \tau_q^{(4)}[\rho_q]\!=\! (3 \pi^2)^{-2/3} \frac{\rho_q^{1/3}}{4320} \left\{ 
          24 \frac{\Delta^2 \rho_q}{\rho_q} 
        - 60 \frac{\vec{\nabla} \rho_q \!\cdot\! \vec{\nabla} (\Delta \rho_q)}
                                                                    {\rho_q^2}
        - 28 \left( \frac{\Delta \rho_q}{\rho_q} \right)^{\!\! 2} 
        - 14 \frac{\Delta (\vec{\nabla} \rho_q)^2}{\rho_q^2} 
                                                           \right.  \nonumber\\
&&  \;\;\;\;\;  \left. 
        + \frac{280}{3} \frac{(\vec{\nabla} \rho_q)^2 \Delta \rho_q}{\rho_q^3}
        + \frac{184}{3} \frac{\vec{\nabla} \rho_q \!\cdot\! \vec{\nabla} 
                                             (\vec{\nabla} \rho_q)^2}{\rho_q^3}
        - 96 \left(\! \frac{\vec{\nabla} \rho_q}{\rho_q} \right)^{\!\!\! 4} \! 
        - 36 \frac{\Delta^2 f_q}{f_q} 
        + 36 \frac{\Delta (\vec{\nabla} f_q)^2}{f_q^2}
                                                           \right.  \nonumber\\
&&  \;\;\;\;\;  \left. 
        - 18 \left( \frac{\Delta f_q}{f_q} \right)^{\!\! 2} 
        - 72 \frac{\vec{\nabla} f_q \!\cdot\! \vec{\nabla} (\vec{\nabla} f_q)^2}
                                                                         {f_q^3}
        + 54 \left( \frac{\vec{\nabla} f_q}{f_q} \right)^{\!\!\! 4}
        + 12 \frac{\Delta (\vec{\nabla} f_q \!\cdot\! \vec{\nabla} \rho_q)}
                                                                {f_q \, \rho_q}
                                                           \right.  \nonumber\\
&&  \;\;\;\;\;  \left. 
        + 24 \frac{\vec{\nabla} f_q \!\cdot\! \vec{\nabla} (\Delta \rho_q)}
                                                                {f_q \, \rho_q}
        - 36 \frac{\vec{\nabla} \rho_q \!\cdot\! \vec{\nabla} (\Delta f_q)}
                                                                {f_q \, \rho_q}
        + 24 \frac{\vec{\nabla} \rho_q \!\cdot\! \vec{\nabla} 
                                         (\vec{\nabla} f_q)^2}{f_q^2 \, \rho_q}
        - 12 \frac{(\vec{\nabla} f_q \!\cdot\! \vec{\nabla} \rho_q) \Delta f_q} 
                                                              {f_q^2 \, \rho_q}
                                                           \right.  \nonumber\\
\end{eqnarray*}
\begin{eqnarray}
&&  \;\;\;\;\;  \left. 
        - 24 \frac{\vec{\nabla} f_q \!\cdot\! \vec{\nabla} 
                   (\vec{\nabla} f_q \!\cdot\! \vec{\nabla} \rho_q)}
                                                              {f_q^2 \, \rho_q}
        - 44 \frac{\vec{\nabla} \rho_q \!\cdot\! \vec{\nabla} 
             (\vec{\nabla} f_q \!\cdot\! \vec{\nabla} \rho_q)}{f_q \, \rho_q^2}
        - 16 \frac{\vec{\nabla} f_q \!\cdot\! \vec{\nabla} 
                                      (\vec{\nabla} \rho_q)^2}{f_q \, \rho_q^2}
                                                           \right.  \nonumber\\
&&  \;\;\;\;\;  \left. 
        - 52 \frac{(\vec{\nabla} f_q \!\cdot\! \vec{\nabla} \rho_q) 
                                                \Delta \rho_q}{f_q \, \rho_q^2} 
        + 30 \frac{(\vec{\nabla} f_q \!\cdot\! \vec{\nabla} \rho_q)^2}
                                                            {f_q^2 \, \rho_q^2}
        + \frac{260}{3} \frac{(\vec{\nabla} f_q \!\cdot\! \vec{\nabla} \rho_q) 
                        (\vec{\nabla} \rho_q)^2}{f_q \, \rho_q^3} \right\} \; .
\label{c2.16}\end{eqnarray}
The interested reader, who might want to use the semiclassical functionals 
calculated here, will find in the Appendix the expression that the 4$^{\rm th}$ 
order kinetic energy density takes in the case of spherical symmetry as well 
as all the other semiclassical functionals developed below.
                                                                    \\[ -0.8ex]

Until now we have not taken into account the spin-orbit interaction. Its 
influence on the semiclassical ETF functionals is treated in ref. 
\cite{GV80} and its contribution $\tau_q^{(4)_{{}_{\rm so}}}$ constitutes 
simply an additive term to the spin-orbit independent part of the kinetic 
energy density considered above. 
According to \cite{GV80}~: 
\begin{eqnarray}
&&  \!\!\!\!\!\!\!\!\!\!
   \tau_q^{(4)_{{}_{\rm so}}}[\rho] \!=\! (3 \pi^{2})^{-2/3} \, 
        \left( \frac{2m}{\hbar^2} \right)^2 \frac{\rho_q^{1/3}}{4 f_q^2} 
         \left\{ \left[ \frac{1}{4} \vec{W}_q \!\cdot\! \Delta \vec{W}_q 
         + \frac{1}{2} \vec{W}_q \!\cdot\! \vec{\nabla} (\mathrm{div}\vec{W}_q)
         + \frac{1}{8} \Delta(\vec{W}_q^2) 
                                                   \right. \right.  \nonumber\\
&& \;\;\;   \left. \left. 
         + \frac{1}{4} (\mathrm{div} \, \vec{W}_q)^2 \right] 
     - \frac{1}{2f_q} \left[ 
       2 \vec{W}_q \cdot (\vec{\nabla}\!f_q \! \cdot \! \vec{\nabla}) \vec{W}_q 
       + \mathrm{div} \vec{W}_q \, (\vec{\nabla}\!f_q \! \cdot \! \vec{W}_q) 
       + \vec{\nabla}\!f_q \cdot (\vec{W}_q \! \cdot \! \vec{\nabla}) \vec{W}_q 
                                                   \right. \right. \nonumber\\
&& \;\;\;  \left. \left. 
         + \vec{W}_q^2 \Delta f_q
         + \vec{W}_q \!\cdot\! \vec{\nabla} 
                                        (\vec{W}_q \!\cdot\! \vec{\nabla}\!f_q)
         - \frac{1}{2} \vec{\nabla}\!f_q \!\cdot\! \vec{\nabla} (\vec{W}_q^2)
                                                                       \right] 
    + \frac{3}{4 f_q^2} \left[ 
          (\vec{\nabla}\!f_q)^2 \vec{W}_q^2 
        + (\vec{W}_q \! \cdot \! \vec{\nabla}\!f_q)^2 
                                                    \right. \right. \nonumber\\
&& \;\;\;  \left. \left. 
        - (\frac{2m}{\hbar^2})^2 \vec{W}_q^4 \right] 
    + \frac{1}{6 \rho_q} \left[ 
          \vec{W}_q \!\cdot\! (\vec{\nabla}\!\rho_q \!\cdot\! \vec{\nabla}) 
                                                                      \vec{W}_q 
        + (\vec{W}_q \!\cdot\! \vec{\nabla}\!\rho_q) \, \mathrm{div} \vec{W}_q 
                                                    \right] \right. \nonumber\\
&& \;\;\;  \left. 
    - \frac{1}{6 f_q \rho_q} \left[ 
          (\vec{\nabla}\!f_q \!\cdot\! \vec{\nabla}\!\rho_q) \vec{W}_q^2 
        + (\vec{\nabla}\!f_q \!\cdot\! \vec{W}_q)
              (\vec{\nabla}\!\rho_q \!\cdot\! \vec{W}_q) \right] \right\} \; . 
\label{c2.20}\end{eqnarray}

It is now interesting to investigate the relative importance of the different 
contributions in eqs. (\ref{c2.2}), (\ref{c2.14}), (\ref{c2.16}) and 
(\ref{c2.20}) to the kinetic energy density obtained when using the 
selfconsistent semiclassical densities generated by a variational procedure 
restricted to functions of the type of eq. (\ref{c2.3a}) as explained above. 
As can be seen on Fig.\ 2~(a) the Thomas-Fermi contribution to $\tau[\rho]$ 
is largely dominant, at least in the nuclear bulk. Semiclassical corrections 
play, however, a significant role in the nuclear surface with a second-order 
correction which is much larger than the fourth-order term (multiplied for 
better visibility by a factor 10 in Fig.\ 2~(a)). 
The different contributions to the second and fourth 
order functional are given respectively in part (b) and (c) of the figure. We 
show the contributions coming form gradient terms of $\rho$ (term 1 and 2 in 
eq.\ (\ref{c2.14})), of $f$ (terms 4 and 5) and the mixed term (term 3) 
as well as the spin-orbit contribution 
(last term) and similarly in part (c) of the figure for the fourth-order term. 
As can be seen, the gradient term of $\rho$ is dominant in 2$^{\rm nd}$ order, 
whereas in 4$^{\rm nd}$ order the spin-orbit contribution becomes also 
crucial.
The selfconsistent HF neutron kinetic energy density is also shown on 
Fig.\ 2~(a). One notices that, except for quantum oscillations in the nuclear 
interior, the HF kinetic energy density is quite nicely reproduced if the 
semiclassical corrections $\tau_2$ and $\tau_4$ are taken into account. 
                                                                   \\[ -9.0ex]
\begin{center}
\vspace*{0.5cm}
\hspace*{0.2cm}\epsfig{file=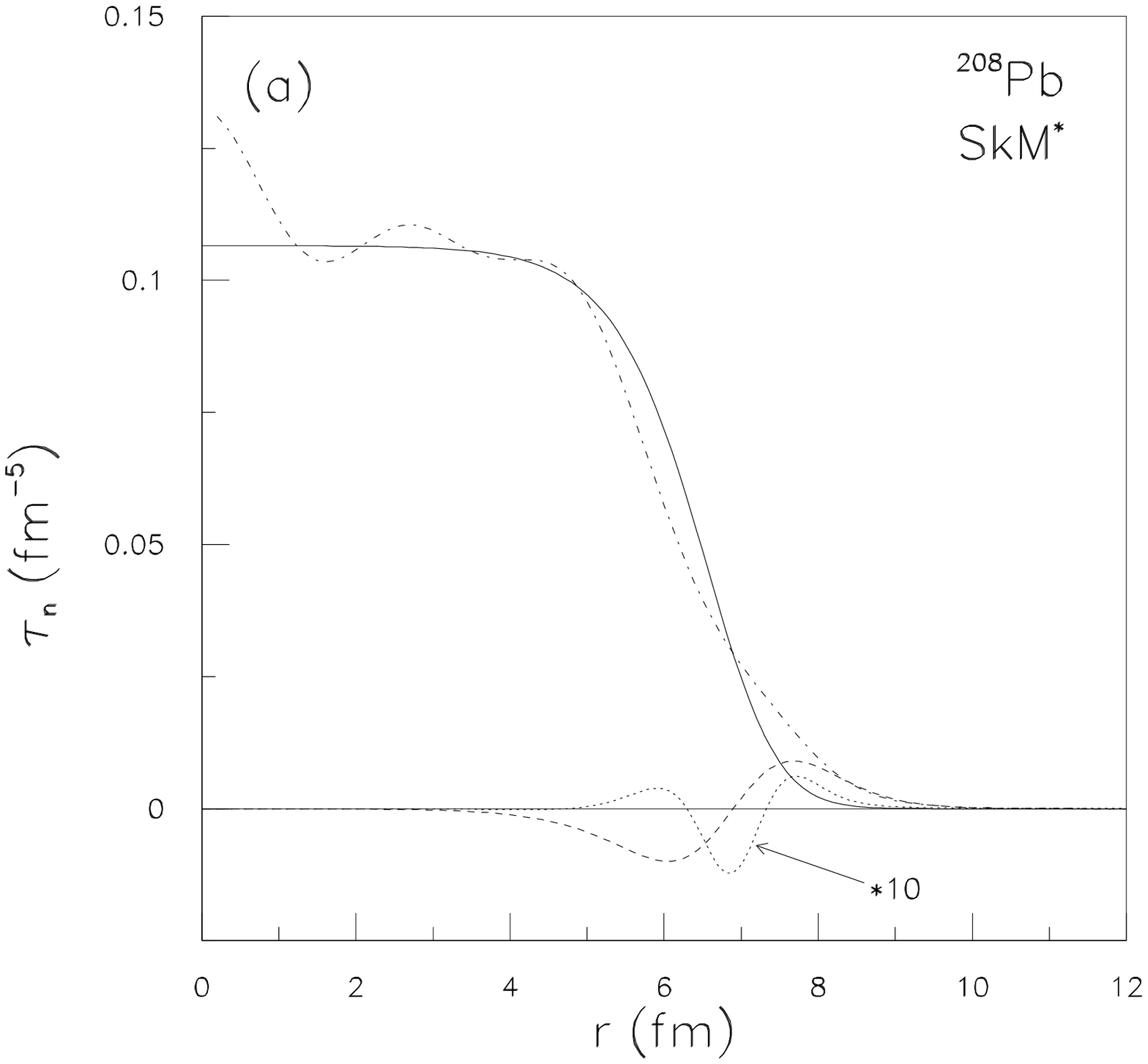,height=6.7cm,width=12.0cm}   \\[ -6.0ex]
\hspace*{0.2cm}\epsfig{file=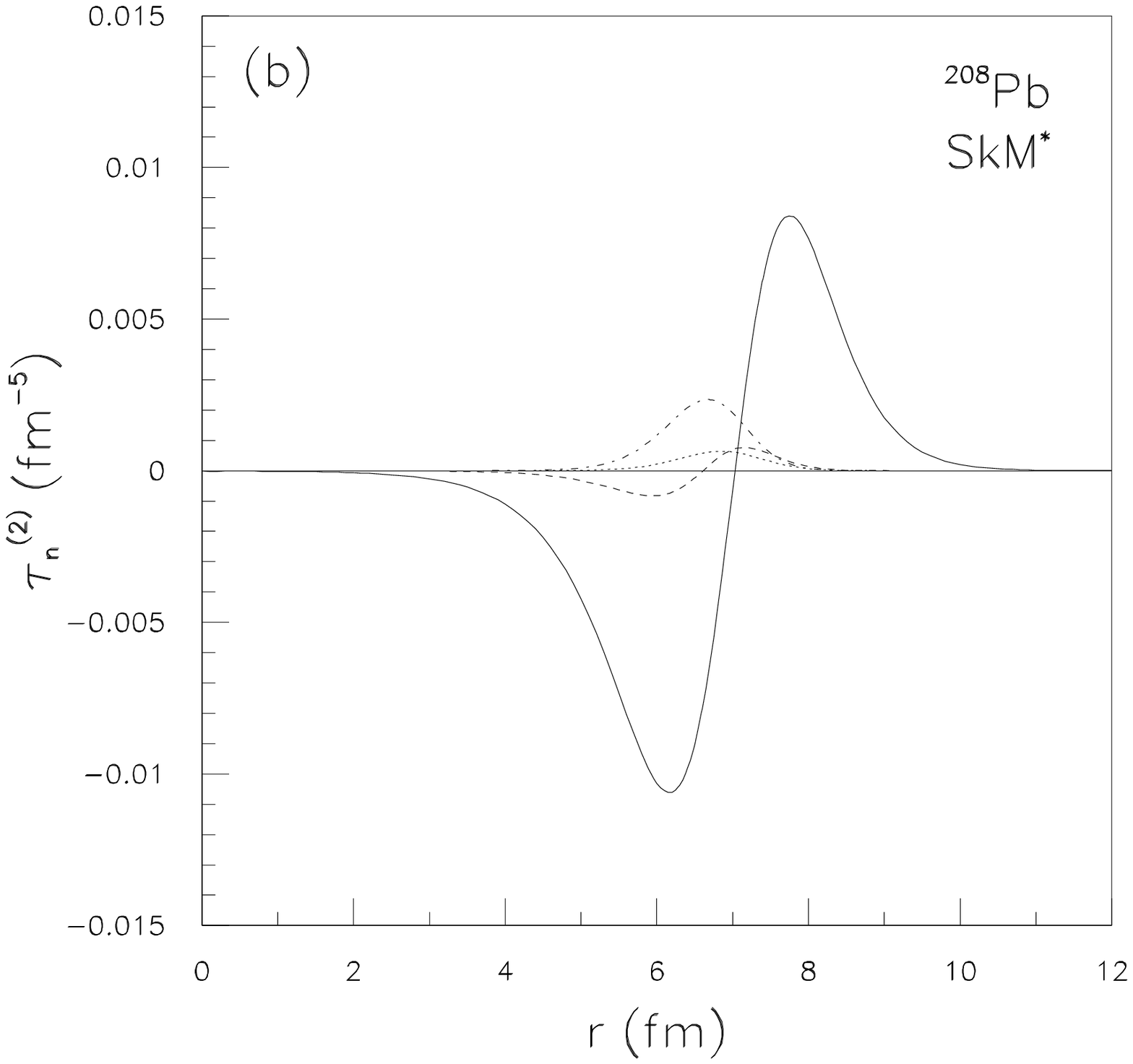,height=6.7cm,width=12.0cm}   \\[ -6.0ex]
\hspace*{0.2cm}\epsfig{file=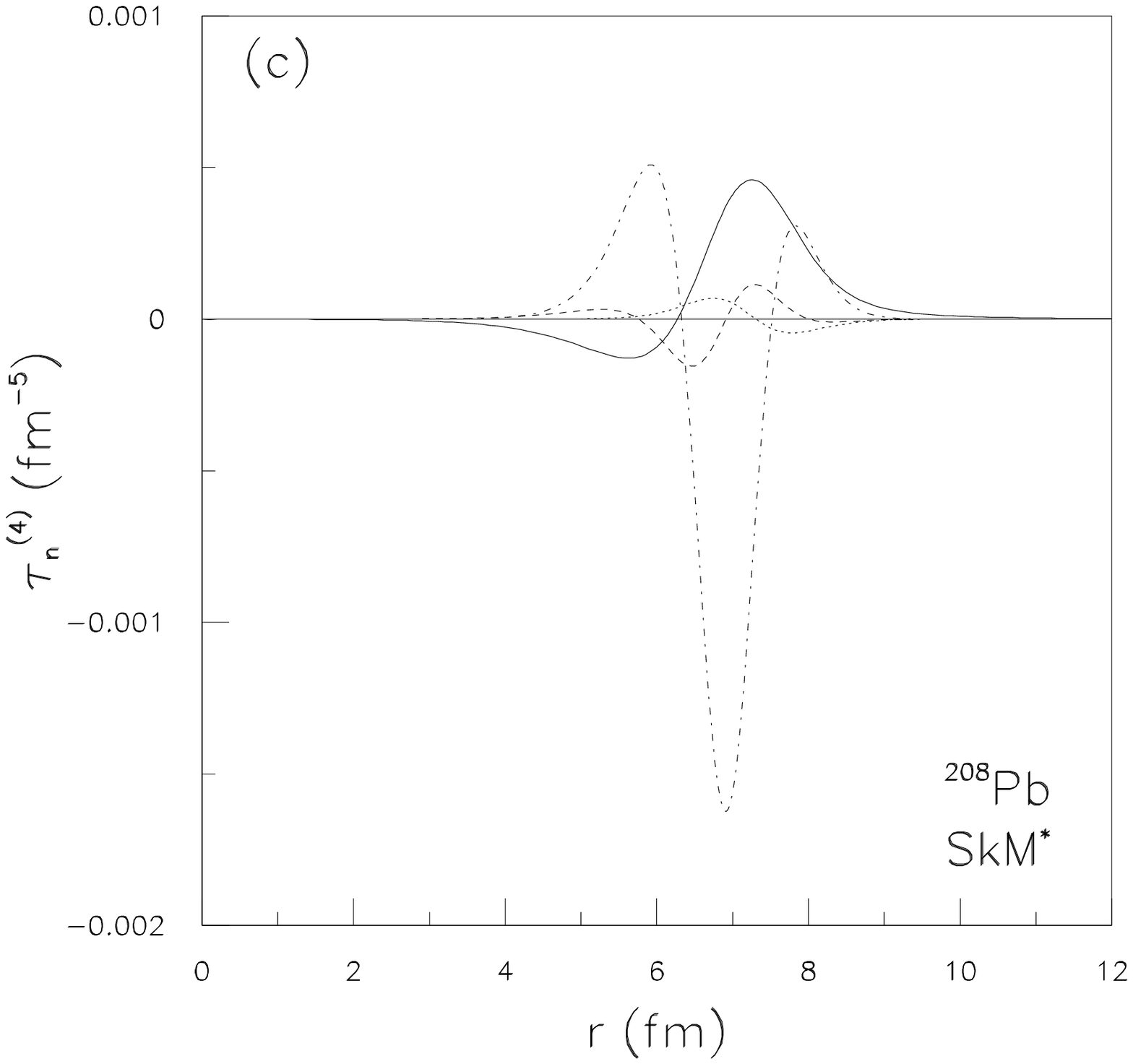,height=6.7cm,width=12.0cm}            
\vspace*{-0.8cm}
\begin{small}
\begin{itemize}
\item[]
Fig. 2 Contributions from the different orders in the semiclassical expansion 
to the kinetic-energy density $\tau[\rho]$ for the selfconsistent neutron 
density distribution shown in Fig. 1 for $^{208}$Pb (TF (solid line), 2$^{nd}$ 
(dashed line) and 4$^{\rm th}$-order multiplied 
(dotted line)) are compared with the 
corresponding HF density (dash-dotted line) (part (a)). Different 
contributions to 2$^{\rm nd}$ (part (b)) and 4$^{\rm th}$ order (part (c)) 
coming from gradient terms of $\rho$ (solid line), of $f$ (dashed), of mixed 
terms containing gradient terms of $\rho$ and $f$ (dotted), and of the 
spin-orbit coupling (dash-dotted line). 
                                                                    \\[  2.5ex]
\end{itemize}
\end{small}
\end{center}

It is interesting to note in this connection that despite the fact that the 
2$^{\rm nd}$ -order contribution $\tau_q^{(2)}(\vec{r})$ is one order of 
magnitude larger than the 4$^{\rm th}$-order contribution, after integration 
the 2$^{\rm nd}$ -order contribution of $\tau[\rho]$ to the total energy, i.e. 
the integral $\sum_q \int f_q(\vec{r}) \tau_q^{(2)}(\vec{r}) d^3r$ is of the 
same order of magnitude than the corresponding 4$^{\rm th}$-order contribution 
(see e.g. \cite{GB80,BGH85}), which seems to indicate that there is a stronger 
cancellation taking place in the 2$^{\rm nd}$-order than in 4$^{\rm th}$-order 
contribution.

We have done the same study for the proton distribution, for other nuclei and 
used other effective interactions of the Skyrme type, namely the Skyrme SIII
force \cite{BFN75} and the SLy4 Skyrme force \cite{CBH97}. The conclusions 
made above remain valid in all these cases, only the relative importance of 
the semiclassical corrections $\tau^{(2)}[\rho]$ and $\tau^{(4)}[\rho]$
increases slightly when one goes from heavy to light nuclei.

As already discussed above, ETF functionals like $\tau^{(ETF)}[\rho]$, 
eq.\ (\ref{c2.13}), up to order $\hbar^4$ constitute the converging part of an 
asymptotic expansion which needs to be truncated. Comparing, indeed, the $\rho$ 
dependence of the different orders of the semiclassical functional 
($\rho^{5/3}$ for $\tau^{(TF)}[\rho]$, $\rho$ for $\tau^{(2)}[\rho]$ and 
$\rho^{1/3}$ for $\tau^{(4)}[\rho]$) one concludes that a term 
$\tau^{(6)}[\rho]$ in the ETF functional would show a $\rho$ dependence of 
the form $\rho^{-1/3}$ and would therefore diverge in the limit 
$r \rightarrow \infty$ for densities that fall off exponentially at large 
distances. 
                                                                    \\[ -0.8ex]

Let us now turn to the spin-orbit density $\vec{J}$. It is given in ref. 
\cite{GV80} in the form of second rank tensor which is related to the 
components of the vector $\vec{J}$ by the relation
$$ J_{\lambda} = \sum_{\mu \nu} \epsilon_{\lambda \mu \nu} \, J_{\mu \nu}    $$
where $\epsilon_{\lambda \mu \nu}$ is the Levi-Civita symbol. The spin 
being a purely quantal property with no classical analogon, there is no 
contribution to the semiclassical functional of $\vec{J}$ in lowest order, 
i.e.\ at the level of the Thomas-Fermi approach whereas one obtains for the 
2$^{\rm nd}$ and 4$^{\rm th}$-order contributions to the semiclassical 
expansion of the spin-orbit density 
\begin{equation}
  \vec{J}_q^{(2)} = - \frac{2 m}{\hbar^2} \, \frac{\rho_q}{f_q} \, \vec{W}_q
\label{c2.27}\end{equation}
and 
\begin{eqnarray}
&&  \!\!\!\!\!\!\!\!\!\! 
  \vec{J}_q^{(4)} = (3 \pi^{2})^{-2/3} \; \frac{2 m}{\hbar^2} \, 
                           \frac{\rho_q^{1/3}}{8 f_q} \left\{ \! - \! \left[ 
                \Delta \vec{W}_q 
              + \vec{\nabla}(\mathrm{div} \vec{W}_q) \right] 
           + \frac{1}{f_q} \! \left[ \vec{W}_q \, \Delta f_q 
              + (\vec{W}_q \!\cdot\! \vec{\nabla}) \vec{\nabla} \! f_q 
                                                   \right. \right.  \nonumber\\
&& \;  \left. \left. 
              + (\vec{\nabla} \! f_q \!\times\! \mathrm{rot} \vec{W}_q) 
              + 2 (\vec{\nabla} \! f_q \!\cdot\! \vec{\nabla}) \vec{W}_q \right]
           - \frac{1}{f_q^2} \left[ (\vec{\nabla} \! f_q)^2 \vec{W}_q 
              + (\vec{\nabla} \! f_q \!\cdot\! \vec{W}_q) \vec{\nabla} \! f_q 
              - 2 \, (\frac{2 m}{\hbar^2})^2 \, \vec{W}_q^3  \right] 
                                                          \right.  \nonumber\\
&& \;\;\;  \left. 
           - \frac{1}{3 \rho_q} 
                \left[ (\vec{\nabla} \! \rho_q \!\cdot\! \vec{\nabla}) \vec{W}_q
              + \mathrm{div} \vec{W}_q \, \vec{\nabla} \! \rho_q 
           - \frac{1}{f_q} \left( 
                (\vec{\nabla} \! f_q \!\cdot\! \vec{\nabla}\!\rho_q) \vec{W}_q
              + (\vec{\nabla} \! f_q \!\cdot\! \vec{W}_q) \vec{\nabla} \! \rho_q
                                                      \right) \right] \right\} 
\label{c2.28}\end{eqnarray}

Let us again investigate the convergence of the semiclassical expansion 
associated this time with the vector field $\vec{J}_q[\rho]$ and compare it 
with the corresponding HF density. We show in Fig. 3 the contributions to 
$\vec{J}_q[\rho]$ from 2$^{\rm nd}$ and from 4$^{\rm th}$ order as well as 
$\vec{J}_{HF}$.
\begin{center}
\vspace*{-0.2cm}
\hspace*{0.2cm}\epsfig{file=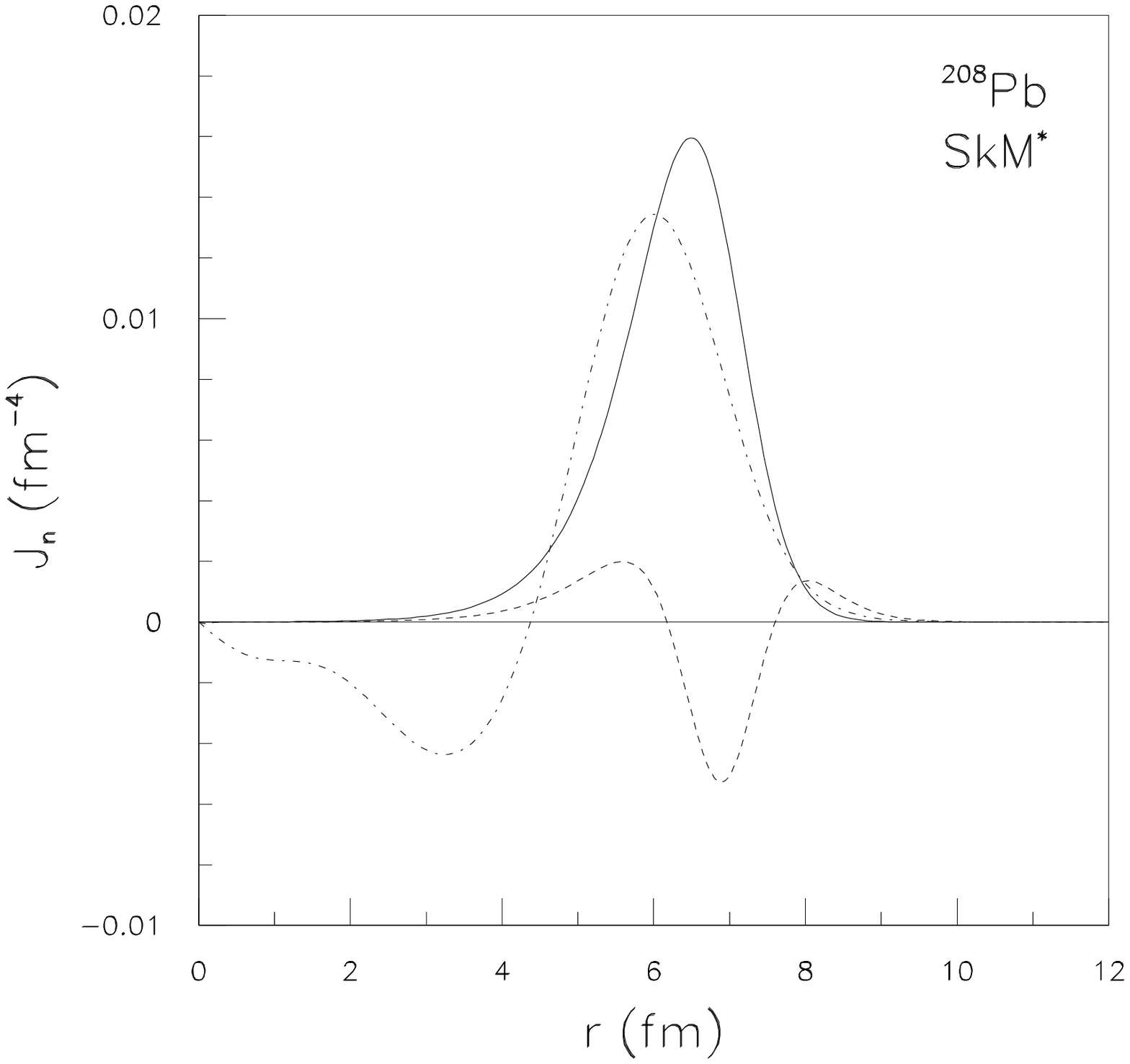,height=7.0cm,width=10.0cm}    
\vspace*{-0.8cm}
\begin{small}
\begin{itemize}
\item[]
Fig. 3. Contributions from the different orders in the semiclassical expansion 
(2$^{\rm nd}$ (solid line) and 4$^{\rm th}$-order (dashed line)) to the radial 
part of the vector field $\vec{J}_q(\vec{r})$ shown here for the selfconsistent 
spherical neutron distribution of $^{208}$Pb as compared to the HF spin-orbit 
density (dash-dotted line). 
                                                                    \\[  2.0ex]
\end{itemize}
\end{small}
\end{center}

We would like to check now that the semiclassical functionals which we have 
written down up to order $\hbar^4$ are indeed correct. We perform this test 
numerically in the following way~:  

One notices that when calculating the total energy through eq. (\ref{c2.12}) 
the kinetic energy density $\tau_q$ does not appear by itself, but only in 
connection with the form factor $f_q$. The $\tau$-dependent part of the total 
energy is simply obtained through the integral $\int d^3r \sum_q f_q \tau_q$. 
In the contribution at order $\hbar^4$ to this integral one can then perform 
integrations by parts to obtain an expression which contains only second order 
derivatives of the density $\rho_q$ and the effective-mass form factor $f_q$ 
\cite{BGH85}.
\begin{eqnarray}
&&  \!\!\!\!\!\!\!\!\!\!
  \int \! f_q \, \tau_q^{(4)} \, d^3r \!=\! (3 \pi^2)^{-2/3} \!
                                      \int \! d^3r \, \rho_q^{1/3} \left\{ \!
       \frac{1}{270} f_q \! \left( \frac{\Delta \rho_q}{\rho_q} 
                                                            \right)^{\!\! 2} \! 
       - \! \frac{1}{240} f_q \, \frac{\Delta \rho_q}{\rho_q} \! \left( 
                         \frac{\vec{\nabla} \rho_q}{\rho_q} \right)^{\!\! 2} \! 
       + \! \frac{1}{810} f_q \! \left( \frac{\vec{\nabla} \rho_q}{\rho_q} 
                                                               \right)^{\!\! 2} 
                                                           \right.  \nonumber\\
&&  \;\;  \left. 
       - \frac{1}{240} \frac{(\Delta f_q)^2}{f_q} 
       + \frac{1}{120} \frac{\Delta f_q \, (\vec{\nabla} f_q)^2}{f_q^2} 
       - \frac{1}{240} \frac{(\vec{\nabla} f_q)^4}{f_q^3}
       + \frac{1}{360} \Delta f_q \, 
            \frac{(\vec{\nabla} f_q \!\cdot\! \vec{\nabla} \rho_q)}{f_q \rho_q}
                                                           \right.  \nonumber\\
&&  \;\;  \left. 
       - \frac{1}{360} \Delta \rho_q \, \frac{(\vec{\nabla} f_q)^2}
                                                                {f_q \, \rho_q}
       - \frac{7}{2160} \Delta f_q \! \left(\! \frac{\vec{\nabla} \rho_q} 
                                                   {\rho_q} \right)^{\!\! 2} \! 
       + \! \frac{1}{540} \! \left(\! \frac{\vec{\nabla} \rho_q}{\rho_q} 
                              \right)^{\!\! 2} \frac{(\vec{\nabla} f_q)^2}{f_q}
       + \frac{7}{2160} \frac{(\vec{\nabla} f_q \!\cdot\! \vec{\nabla} \rho_q)}
                                                              {f_q \, \rho_q^2}
                                                           \right.  \nonumber\\
&&  \;\;  \left. 
       - \frac{11}{3240} \! \left( \frac{\vec{\nabla} \rho_q}{\rho_q} 
                                                              \right)^{\!\! 2}
                \frac{(\vec{\nabla} f_q \!\cdot\! \vec{\nabla} \rho_q)}{\rho_q}
       + \frac{7}{1080} \, \Delta \rho_q \, 
              \frac{(\vec{\nabla} f_q \!\cdot\! \vec{\nabla} \rho_q)}{\rho_q^2}
       + \frac{1}{180} \, \Delta f_q \, \frac{\Delta \rho_q}{\rho_q} \right\} 
                                                                          \; ,
                                                          \nonumber\\[ -1.0ex]
\label{c2.17}\end{eqnarray}
where $\tau_q^{(4)}$ is the spin-independent part of the kinetic energy density.

The dependence of the total energy density on the spin degrees of freedom 
enters in two different ways~: through the spin-orbit part of the kinetic 
energy density $\tau_q^{so}$ and through a term $\vec{J}_q \!\cdot\! \vec{W}_q$.
This total spin-dependence is then given by 
$ \sum_q \int \left( \frac{\hbar^2}{2 m} f_q \tau_q^{so}[\rho] 
            + \vec{W}_q \cdot \vec{J}_q[\rho] \right) d^3r  $. 
It can be shown after some integration by parts with eqs.\ (\ref{c2.14}), 
(\ref{c2.20}), (\ref{c2.27}) and (\ref{c2.28}) that in the different orders 
of the semiclassical expansion this integral takes on the simple form 
\cite{BGH85} 
\begin{equation}
   \int \left[ \frac{\hbar^2}{2 m} f_q \, \tau_q^{(2)_{{}_{so}}}[\rho]
              + \vec{W}_q \cdot \vec{J}_q^{(2)}[\rho] \right] d^3r =  
             - \frac{m}{\hbar^2} \sum_q \, 
                 \int \frac{\rho_q}{f_q} \, \vec{W}_q^2 d^3r
\label{c2.31}\end{equation}
                                                                    \\[ -2.0ex]
and that 
\begin{eqnarray}
&&  \!\!\!\!\!\!\!\!\!\!\!\!\!\!\!\!\!\!\!\!\!\!\!\!\!\!\!\!\!\!\!\!\!\!\!
   \int \left[ \frac{\hbar^2}{2 m} f_q \tau_q^{(4)_{{}_{so}}}[\rho]
              + \vec{W}_q \cdot \vec{J}_q^{(4)}[\rho] \right] d^3r =  
                                                                   \nonumber\\ 
&& \;\;\;  
     = (3 \pi^{2})^{-2/3} \, \frac{m}{\hbar^2} \; \int \frac{\rho_q^{1/3}}{f_q}
         \left\{ \frac{1}{4} (\mathrm{div} \, \vec{W}_q)^2 
             - \frac{3}{8} \, \mathrm{div} \, \vec{W}_q \, 
                          \frac{(\vec{\nabla}\!f_q \! \cdot \! \vec{W}_q)}{f_q}
                                                           \right. \nonumber\\ 
&& \;\;\;\;\;\;\;  \left.
             + \frac{1}{16} \, \vec{W}_q^2 \, \frac{\Delta f_q}{f_q} 
             + \frac{1}{8} \frac{(\vec{\nabla}\!f_q \! \cdot \! \vec{W}_q)^2}
                                                                        {f_q^2}
             + \frac{1}{24} \, \mathrm{div} \, \vec{W}_q \, 
                    \frac{(\vec{\nabla}\!\rho_q \! \cdot \! \vec{W}_q)}{\rho_q}
                                                            \right. \nonumber\\
&& \;\;\;\;\;\;\;  \left.
             + \frac{1}{48} \, \vec{W}_q^2 \, \frac{\Delta \rho_q}{\rho_q} 
             - \frac{1}{72} \, \vec{W}_q^2 
                     \left( \frac{\vec{\nabla} \rho_q}{\rho_q} \right)^{\!\!2} 
             + \frac{1}{2} \, \left( \frac{m}{\hbar^2} \right)^2 
                            \frac{\vec{W}_q^4}{f_q^2} \right\} d^3r \; .
\label{c2.32}\end{eqnarray}
                                                                    \\[ -1.0ex]

We have tested the semiclassical functionals given above by verifying 
numerically that these integral relations (\ref{c2.17}), (\ref{c2.31}) 
and (\ref{c2.32}) hold true.

We have also evaluated different integrals involving 
these functionals and which have been calculated in ref. \cite{CPV90}. 
We obtain agreement with their results of the order of 1 to 2 \%, which is 
of the same order as their agreement between the results of a full variational 
calculation and one in the restricted subspace of modified Fermi functions.
                                                                    \\[ -2.0ex]

As can be seen on eq. (\ref{c2.8}) only the divergence of the vector field 
$\vec{J}$ is present in the expression of the central one-body potential. 
One obtains from eqs.\ (\ref{c2.27}) and (\ref{c2.28}) respectively the 
contributions to 2$^{\rm nd}$ order 
\begin{equation}
  \mathrm{div} \vec{J}_q^{(2)} = - \frac{2 m}{\hbar^2} \frac{1}{f_q} 
                             \left[ \rho_q \mathrm{div} \vec{W}_q
                                   + (\vec{\nabla} \rho_q \!\cdot\! \vec{W}_q) 
            - \frac{\rho_q}{f_q} (\vec{\nabla} f_q \!\cdot\! \vec{W}_q) \right]
\label{c2.33}\end{equation}
and, after some lengthy but straightforward calculation, 
to $4^{\mathrm{th}}$ order 
\begin{eqnarray}
&&  \!\!\!\!\!\!\!\!\!\! 
  \mathrm{div} \vec{J}_q^{(4)} = (3 \pi^{2})^{-2/3} \; \frac{2 m}{\hbar^2} \, 
           \frac{\rho_q^{1/3}}{8 f_q} \left\{\!
     - 2 \mathrm{div} \left[ \vec{\nabla} (\mathrm{div} \vec{W}_q) \right] 
     + \frac{1}{2f_q} \left[ 2 \Delta f_q \mathrm{div} \vec{W}_q 
        + (\vec{\nabla}^3 \! f_q \!\cdot\! \vec{W}_q)
                                                   \right. \right.  \nonumber\\
&& \left. \left. \;\;\;\;\; 
        + 3 \Delta (\vec{\nabla} \! f_q \!\cdot\! \vec{W}_q) 
        + (\vec{\nabla} \! f_q \!\cdot\! \Delta \vec{W}_q)
        + 4 \vec{\nabla} \! f_q \!\cdot\! \vec{\nabla} (\mathrm{div} \vec{W}_q) 
                                                                        \right] 
     - \frac{1}{f_q^2} \left[ (\vec{\nabla} \! f_q)^2 \mathrm{div} \vec{W}_q 
                                                   \right. \right.  \nonumber\\
&& \left. \left.  \;\;\;\;\;  
        + 3 \Delta \! f_q (\vec{\nabla} \! f_q \!\cdot\! \vec{W}_q) 
        + 5 \vec{\nabla} \! f_q \!\cdot\! 
                 (\vec{W}_q \!\cdot\! \vec{\nabla}) \vec{\nabla} \! f_q 
        + 5 \vec{\nabla} \! f_q \!\cdot\! 
                      (\vec{\nabla} \! f_q \!\cdot\! \vec{\nabla}) \vec{W}_q) 
                                                   \right. \right.  \nonumber\\
&& \left. \left.  \;\;\;\;\;  
        + 2 \vec{\nabla} \! f_q \!\cdot\! 
                        (\vec{\nabla} \! f_q \!\times\! \mathrm{rot} \vec{W}_q) 
        - 2 \,(\frac{2 m}{\hbar^2})^2 \,\left(\vec{W}_q^2 \mathrm{div}\vec{W}_q 
                       + \vec{W}_q \!\cdot\! \vec{\nabla} (\vec{W}_q^2) \right)
                                                   \right] \right.  \nonumber\\
&& \left.  
     + \frac{6}{f_q^3} \left[ (\vec{\nabla} \! f_q)^2 
                                      (\vec{\nabla} \! f_q \!\cdot\! \vec{W}_q) 
        - (\frac{2 m}{\hbar^2})^2 \, \vec{W}_q^2 
                                     (\vec{\nabla} \! f_q \!\cdot\! \vec{W}_q) 
                                                   \right] 
     - \frac{1}{6 \rho_q} \left[ 
                            \Delta (\vec{\nabla} \! \rho_q \!\cdot\! \vec{W}_q)
        + (\vec{\nabla} \! \rho_q \!\cdot\! \Delta \vec{W}_q)
                                                   \right. \right.  \nonumber\\
&& \left. \left.  \;\;\;\;\;  
        - (\vec{\nabla}^3 \! \rho_q \!\cdot\! \vec{W}_q)
        + 2 \Delta \rho_q \, \mathrm{div} \vec{W}_q 
        + 6 \vec{\nabla} \! \rho_q \!\cdot\! \vec{\nabla} 
                                               (\mathrm{div} \vec{W}_q) \right] 
     + \frac{1}{3 f_q \rho_q} \left[ 2 (\vec{\nabla} \! f_q \!\cdot\! 
                                    \vec{\nabla} \rho_q) \mathrm{div} \vec{W}_q
                                                  \right.  \right.  \nonumber\\
&& \left. \left.  \;\;\;\;\;  
        + \vec{W}_q \!\cdot\! \vec{\nabla} 
                            (\vec{\nabla} \! f_q \!\cdot\! \vec{\nabla} \rho_q) 
        + \Delta \rho_q (\vec{\nabla} \! f_q \!\cdot\! \vec{W}_q) 
        + 2 \vec{\nabla} \! \rho_q \!\cdot\! \vec{\nabla} \! 
                                     (\vec{\nabla} \! f_q \!\cdot\! \vec{W}_q) 
        + \Delta f_q (\vec{\nabla} \! \rho_q  \!\cdot\! \vec{W}_q) 
                                                   \right. \right.  \nonumber\\
&& \left. \left.  \;\;\;\;\;  
        + \vec{\nabla} \! \rho_q \!\cdot\! 
                       (\vec{\nabla} \! f_q \!\times\! \mathrm{rot} \vec{W}_q) 
        + 2 \vec{\nabla} \! \rho_q \!\cdot\! 
                        (\vec{\nabla} \! f_q \!\cdot\! \vec{\nabla}) \vec{W}_q) 
                                                                        \right] 
     - \frac{1}{3 f_q^2 \rho_q} \left[ 5 (\vec{\nabla} \! f_q \!\cdot\! 
                 \vec{\nabla} \rho_q) (\vec{\nabla} \! f_q \!\cdot\! \vec{W}_q) 
                                                   \right. \right.  \nonumber\\
&& \left. \left.  \;\;\;\;\;  
        + (\vec{\nabla} \! f_q)^2 (\vec{\nabla} \! \rho_q \!\cdot\! \vec{W}_q) 
        - 2 (\frac{2 m}{\hbar^2})^2 \, \vec{W}_q^2 
                          (\vec{\nabla} \! \rho_q \!\cdot\! \vec{W}_q) \right]
     + \frac{2}{9 \rho_q^2} \left[ \vec{\nabla} \! \rho_q \!\cdot\! 
                    (\vec{\nabla} \! \rho_q \!\cdot\! \vec{\nabla}) \vec{W}_q) 
                                                   \right. \right. \nonumber\\
&& \left. \left.  \;\;\;\;\;  
     + (\vec{\nabla} \! \rho_q)^2 \mathrm{div} \vec{W}_q \right] 
     - \frac{2}{9 f_q \rho_q^2} 
                     \left[ (\vec{\nabla} \! f_q \!\cdot\! \vec{\nabla} \rho_q) 
                                   (\vec{\nabla} \! \rho_q \!\cdot\! \vec{W}_q)
        + (\vec{\nabla} \! \rho_q )^2 (\vec{\nabla} \! f_q \!\cdot\! \vec{W}_q) 
                                            \right] \right\} \;\; .  
\label{c2.34}\end{eqnarray}
%
\section{Average nuclear potentials}

One now controls all the ingredients which enter into the calculation of the 
nuclear central potentials, eq. (\ref{c2.8}), for effective interactions of the 
Skyrme type. 
It is now interesting to look at the {\it convergence} of the expressions 
which define these average fields calculated with the semiclassical functionals 
$\tau_q[\rho]$ and $\vec{J}_q[\rho]$ and to check how these potentials compare 
with the ones obtained in the HF approach.

We therefore show in Fig.\ 4 the neutron and proton nuclear central potentials 
for the nucleus $^{208}$Pb obtained for the Skyrme interaction SkM$^*$. The 
Coulomb potential $V_{Coul}$ for the proton field has been left out in this 
investigation, because it is directly given through 
the proton density (see eq.\ (\ref{c2.11})). As the latter 
is very well reproduced, except for quantum oscillations in the nuclear 
interior, we already know that the Coulomb potential calculated through this 
semiclassical density will, indeed, reproduce on the average the exact one 
calculated from the quantum-mechanical densities. 

We show a comparison between the HF neutron and proton central potentials 
with the ones obtained using the selfconsistent semiclassical densities 
$\rho_n$ and $\rho_p$ but restricting ourselves to the TF approximations for 
the functional $\tau[\rho]$, eq.\ (\ref{c2.2}), and $\vec J[\rho]$ (which is 
zero as explained above). We do not want to call this the Thomas-Fermi 
approximation to the nuclear central fields since even if we have used the 
above mentioned functionals in their Thomas-Fermi approximation, the nuclear 
structure has been determined through a full variational calculation including 
the functionals up to order $\hbar^4$.

As seen on the figure the reproduction of the HF selfconsistent fields is 
already quite remarkable at the lowest (TF) order in the semiclassical 
expansion. Apart from shell oscillations in the nuclear interior and a small 
wiggle in the TF potential in the surface region the agreement seems very 
satisfactory. 
\begin{center}
\vspace*{0.02cm}
\hspace*{0.2cm}\epsfig{file=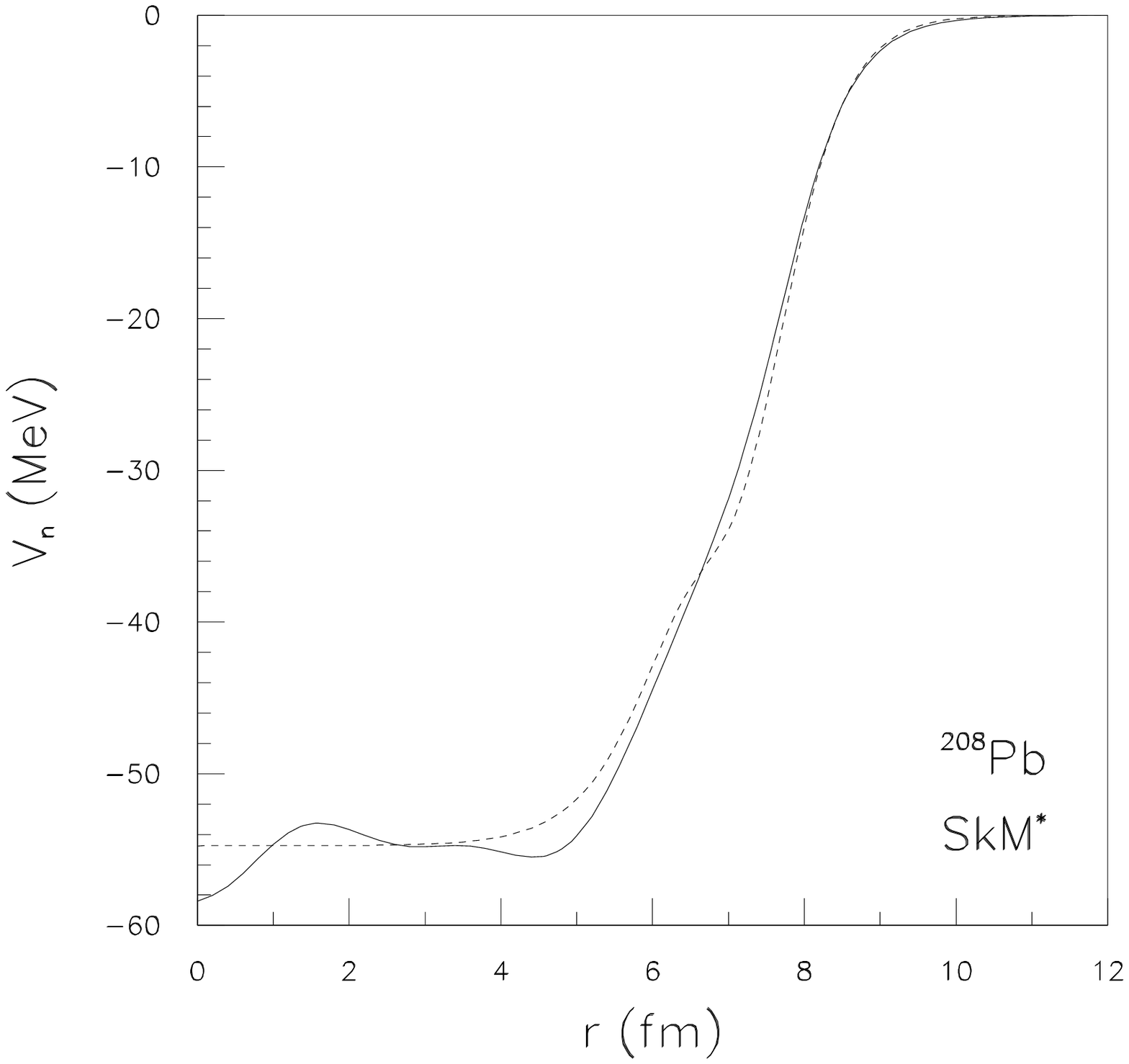,height=6.8cm,width=10.0cm}            
\end{center}
\begin{center}
\vspace*{-1.2cm}
\hspace*{0.2cm}\epsfig{file=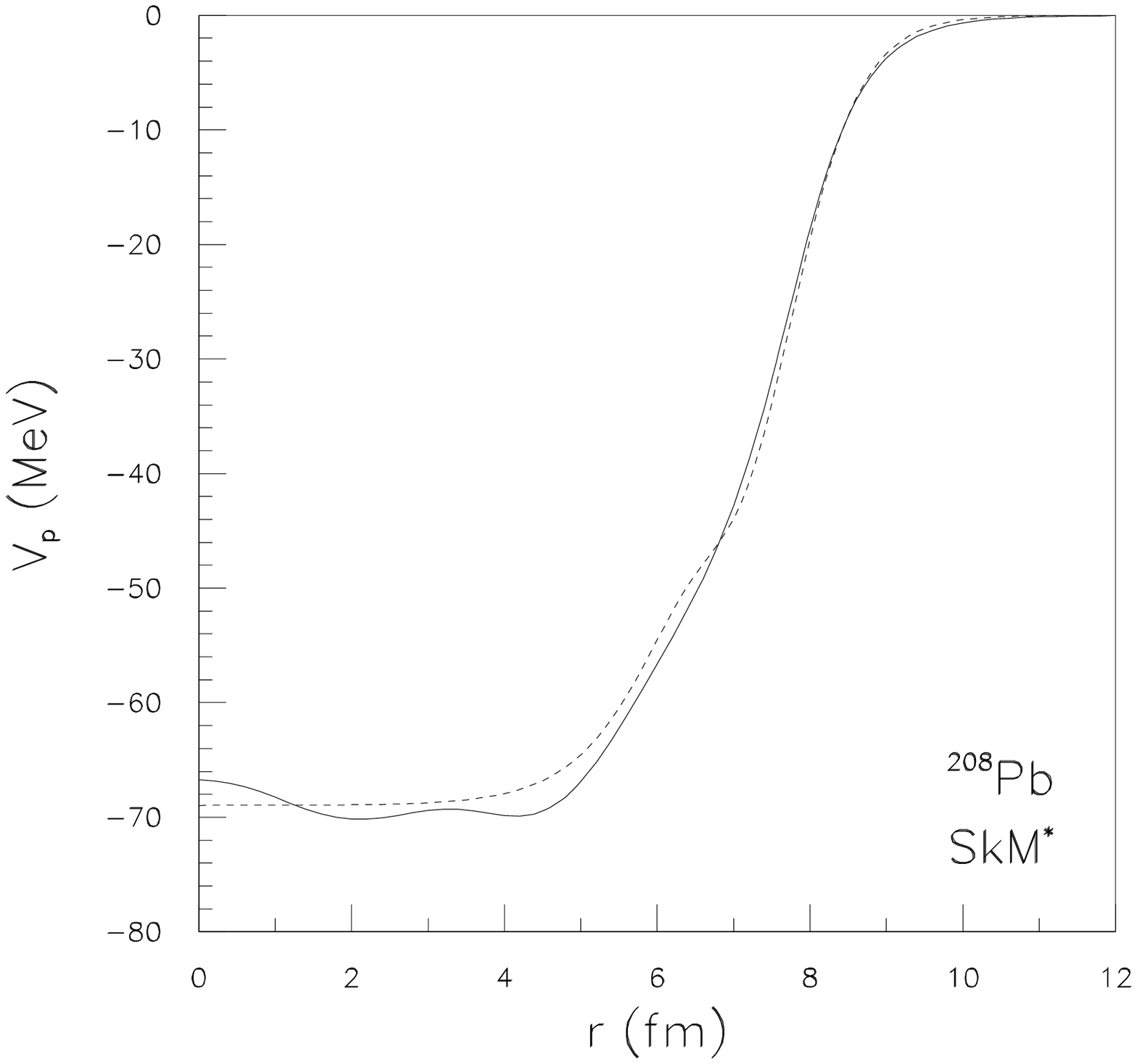,height=7.0cm,width=10.0cm}            
\vspace*{-0.8cm}
\begin{small}
\begin{itemize}
\item[]
Fig.\ 4. Comparison of the Hartree-Fock central nuclear potentials (solid line) 
for protons and neutrons with the corresponding semiclassical potentials 
(dashed line) obtained using the Thomas-Fermi approximation for the 
functionals $\tau[\rho]$ and $\vec{J}[\rho]$. 
                                                                    \\[ -0.5ex]
\end{itemize}
\end{small}
\end{center}

It is now interesting to study the contributions to the nuclear central 
fields coming from higher orders in the semiclassical expansion. For this 
reason we show in Fig.\ 5 the corrections $\delta V_n^{(2)}$ and 
$\delta V_n^{(4)}$ defined as (see eq.\ (\ref{c2.8}))
\[  \delta V_n^{(2)} = (B_3 + B_4) \tau_n^{(2)} + B_3 \tau_p^{(2)} 
  + B_9 (2 \, \mathrm{div} \vec{J}_n^{(2)} + \mathrm{div} \vec{J}_p^{(2)})\;, \]
                                                                    \\[ -8.0ex]
\begin{equation}
  {}
\label{c2.37}\end{equation}
                                                                    \\[ -8.0ex]
\[  \delta V_n^{(4)} = (B_3 + B_4) \tau_n^{(4)} + B_3 \tau_p^{(4)} 
  + B_9 (2 \, \mathrm{div} \vec{J}_n^{(4)} + \mathrm{div} \vec{J}_p^{(4)})\;  \]
                                                                    \\[ -1.5ex]
together with the semiclassical TF potential already shown in Fig.\ 4. 
It turns out that these corrections are rather small and we have to multiply 
$\delta V_n^{(2)}$ by a factor of 10 and $\delta V_n^{(4)}$ by a factor of 100 
to make their relative importance better visible on figure 5.
\begin{center}
\vspace*{-0.2cm}
\hspace*{0.2cm}\epsfig{file=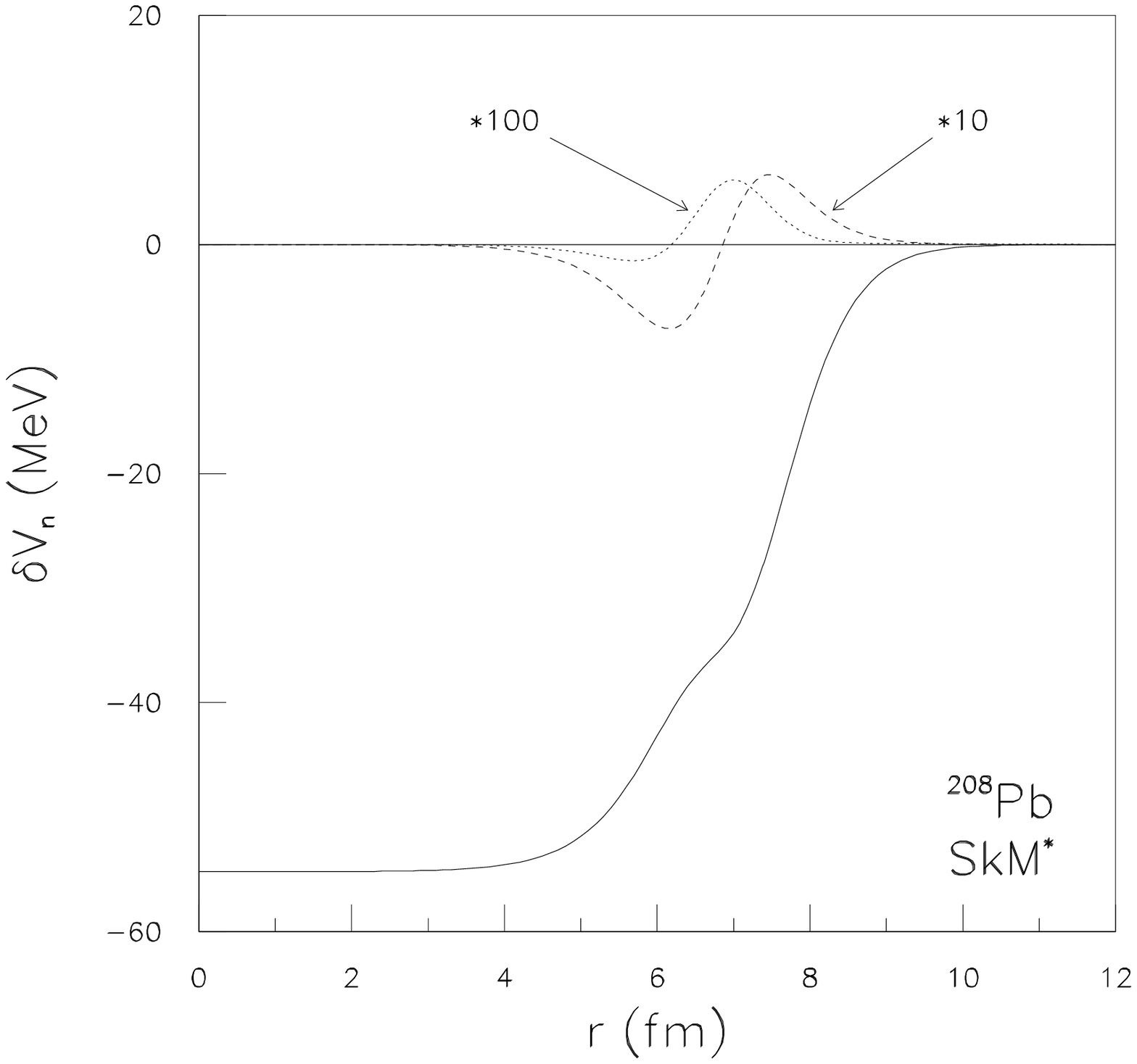,height=7.0cm,width=10.0cm}            
\vspace*{-0.8cm}
\begin{small}
\begin{itemize}
\item[]
Fig. 5. Semiclassical neutron potential using the TF approximation (solid line) 
of the semiclassical functionals $\tau[\rho]$, and $\vec J[\rho]$ and 
corrections coming from second (dashed) and fourth order (dotted line) in the 
semiclassical expansion. For better visibility the second order correction has 
been multiplied by a factor of 10 and the fourth order by a factor of 100.
                                                                    \\[  1.5ex]
\end{itemize}
\end{small}
\end{center}
It can be seen that both these 
terms give a contribution in the nuclear surface where the lowest order 
term showed some deviation from the HF potentials. It is therefore to be 
expected that potential using the second order functionals $\tau^{(2)}[\rho]$ 
and $\vec{J}^{(2)}[\rho]$ will partially 
correct for this deficiency and be quite close 
to the selfconsistent HF potentials. Due to the smallness of the 4$^{\rm th}$ 
order term in Fig.\ 5 we can expect the semiclassical potentials obtained 
using the full semiclassical functionals up to 4$^{\rm th}$ order to be 
practically indistinguishable from the ones using the 2$^{\rm nd}$ order 
corrections only. This conclusion is, indeed, confirmed on Fig.\ 6. 
\begin{center}
\vspace{-0.3cm}
\hspace*{0.2cm}\epsfig{file=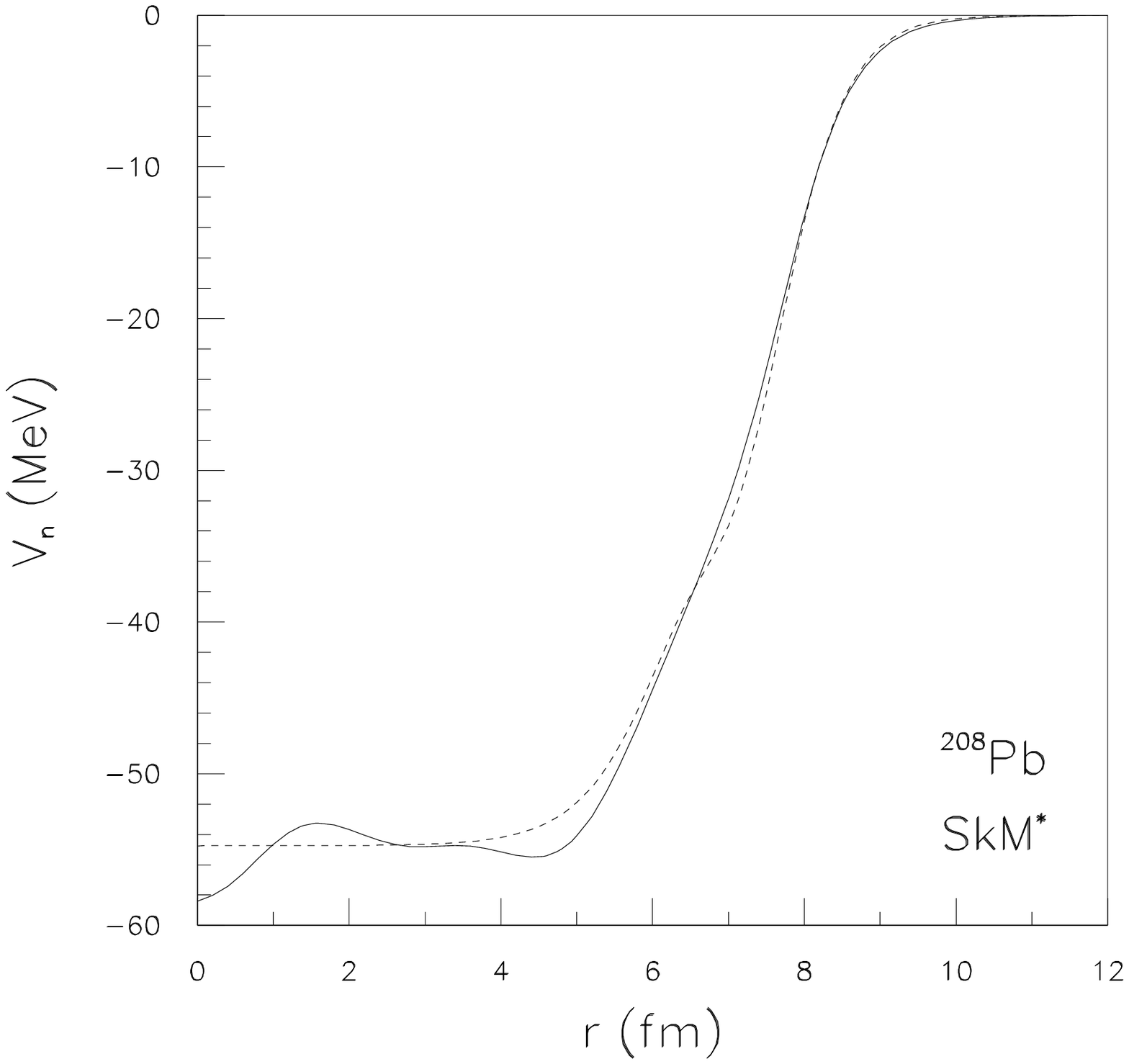,height=8.0cm,width=10.0cm}            
\vspace*{-1.0cm}
\hspace*{0.2cm}\epsfig{file=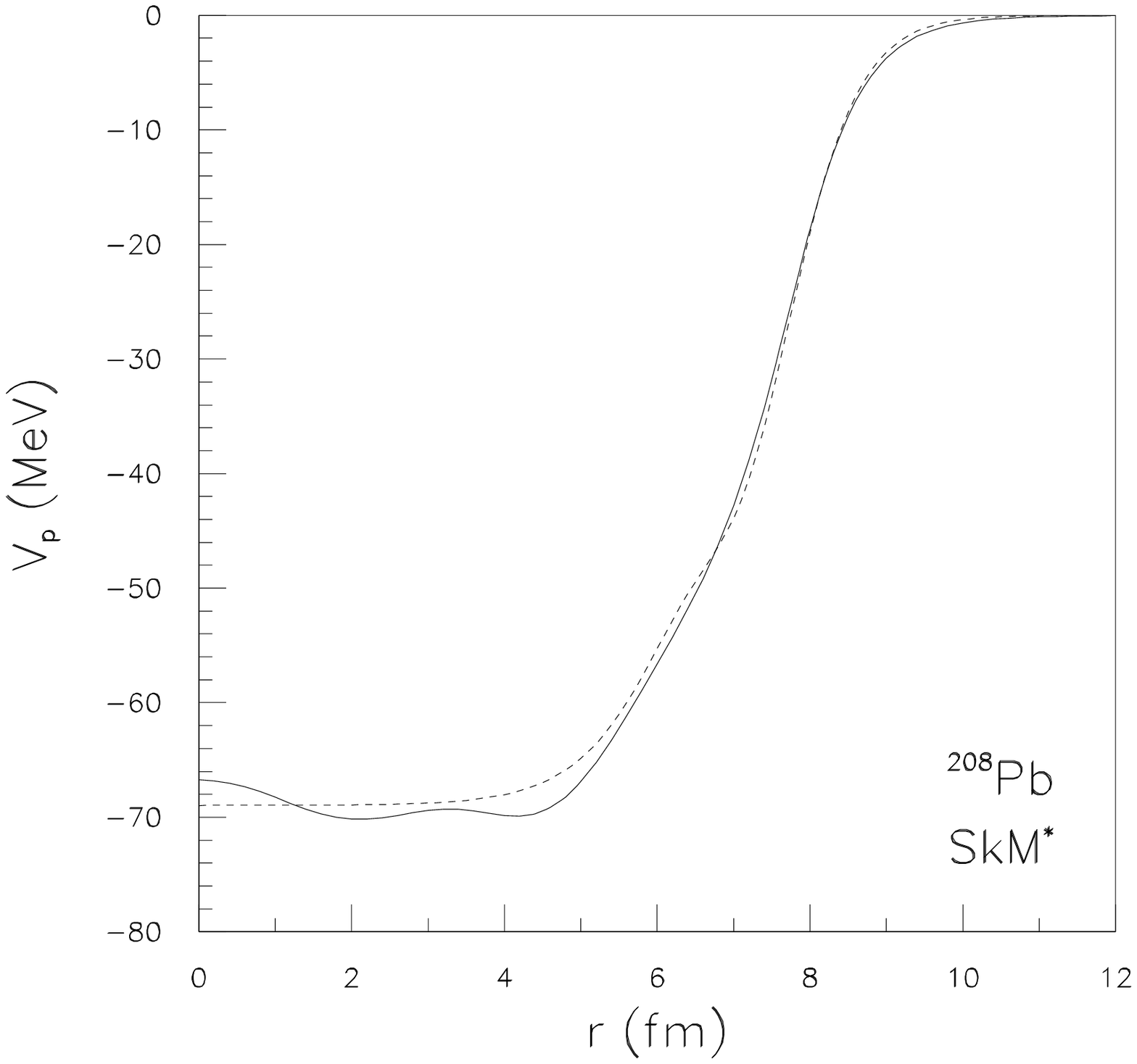,height=8.0cm,width=10.0cm}            
\vspace*{0.2cm}
\begin{small}
\begin{itemize}
\item[]
Fig. 6. Comparison of the Hartree-Fock central nuclear potentials (solid line) 
for protons and neutrons with the semiclassical potentials (dashed line) 
obtained by using the semiclassical functionals $\tau[\rho]$ and $\vec{J}[\rho]$
up to order $\hbar^2$ (dashed line). The semiclassical potentials including the 
functionals up to order $\hbar^4$ are indistinguishable from the latter ones.
                                                                    \\[  1.5ex]
\end{itemize}
\end{small}
\end{center}

The same kind of calculations have been performed also for lighter nuclei 
down to $^{40}$Ca. The quality of the agreement between HF and semiclassical 
potentials is the same as the one obtained for the nucleus $^{208}$Pb studied 
above. As already mentioned, the effective-mass form factor $f_q(\vec r)$, 
eq.\ (\ref{c2.9}) and the spin-orbit potential $\vec W_q(\vec r)$, eq.\ 
(\ref{c2.10}) are, except for shell oscillations in the nuclear interior, 
very well reproduced, since the nuclear densities which directly determine 
these quantities are well reproduced. For this reason we do not explicitly 
show these quantities. These conclusions remain valid when other effective 
interactions of the Skyrme type are used.
\vspace{-0.2cm}
\section{Summary and conclusions }
%
We have demonstrated that using the Extended Thomas-Fermi approach one is not 
only able to give a very precise description of average nuclear properties but 
that this method is also able to reproduce quite nicely local quantities, not 
only neutron and proton density distributions but also the corresponding nuclear
central potentials, effective-mass form factors and spin-orbit potentials. 
These are precisely the ingredients of the Schr\"odinger like 
Hartree-Fock equation, eq.\ (\ref{c2.7}), which arises from the variational 
principle. Within this semiclassical approach which relies on a density 
variational calculation one should therefore be able to solve in an approximate 
way the quantum mechanical problem without having to go through the full 
selfconsistency problem of the HF approach. This is the essential idea of an 
approximate solution of the HF problem known as the ``expectation value method''
(EVM) \cite{Br77}. It consists in constructing the ground state  Slater 
determinant from the eigenfunctions of eq.\ (\ref{c2.7}) using the ETF fields 
$V_q(\vec{r})$, $m_q^*(\vec{r})$ and $\vec{W}_q(\vec{r})$ (to second order in 
the ETF expansions) and calculating with this Slater determinant the 
expectation value of the total Skyrme Hamiltonian. 
The rapid convergence of the ETF functionals demonstrated above explains, 
a posteriori, the success of this approach.

The interested reader might object that nowadays, where computational power 
has been increased tremendously, there is no real need for semiclassical 
approximations, but all calculations of nuclear structure should be directly 
performed at the level of the Hartree-Fock model (or beyond). The point, 
however, is that as soon as one is interested in nuclear systematics where 
one is looking at the behavior of nuclei over a wide range of the nuclear 
chart, semiclassical approximations are without any credible substitute. 

A point that might e.g. be interesting to study is the variation of the 
diffuseness of the nuclear densities and central potentials when increasing 
the nuclear excitation and/or when going to rotating nuclei. 
The approach we have developed here is, indeed, easily generalized to the 
description of excited or rotating nuclei. If one is interested in 
{\it hot} nuclear systems one simply need to replace the semiclassical 
functionals for $\tau[\rho]$ and $\vec{J}[\rho]$ which we have developed by 
those derived in ref. \cite{BBD85} for the ETF approach at finite temperature. 
In this case the coefficients in the semiclassical expansions, eqs. 
(\ref{c2.2}), (\ref{c2.14}), (\ref{c2.16}), (\ref{c2.20}) etc.\ are to be 
replaced by a combination of Fermi integrals \cite{BBD85}. 

If on the other hand one is interested in the description of systems breaking 
time reversal symmetry, the energy density, eq.\ (\ref{c2.6}), will be changed 
\cite{EBG75,BBQ94} and some additional densities will appear, such as the 
current density $\vec j(\vec r)$ or the spin-vector density $\vec \rho(\vec r)$ 
which, in the case of rotations, are a manifestation of the time-odd part 
of the density matrix generated by the cranking piece of the Hamiltonian. 
This causes not only a change in the analytical form of quantities such as 
the average potentials $V_q(\vec r)$ which are now going to depend on these 
additional densities but also leads to the appearance of additional terms in 
the functionals $\tau[\rho]$ and $\vec{J}[\rho]$. These functionals have 
already been determined in \cite{BBQ94} which makes it quite straightforward 
to calculate selfconsistent semiclassical fields such as $V_q(\vec r)$ for 
rotating nuclei very similarly to what we have done here in the static case 
and investigate the dependence of these quantities with increasing angular 
momentum. Investigations along these lines are currently in progress. 

The present method can also be profitably exploited to establish some easy 
to use parametrization (as function of mass number A and isospin parameter 
I=(N-Z)/A) of central and spin-orbit potential and effective mass and this 
over a wide region of the nuclear chart \cite{BPB01}. Such an investigation 
could be the ideal starting point for Strutinski shell correction calculations 
as e.g. formulated in the so-called Extended Thomas-Fermi plus Strutinski 
integral (ETFSI) method (see ref. \cite{APD95} and references given therein) 
where average mean fields like the ones investigated here are used to determine 
shell corrections.

\newpage
\noindent 
{\bf Acknowledgments} 
                                                                    \\[ -1.0ex]

The authors are grateful to K.\ Pomorski for having suggested the 
investigation undertaken here and for his constant interest in this work. 
The authors have also benefitted from fruitful discussions with P.\ Quentin. 
One of us (K.B.) gratefully acknowledges the hospitality extended to him 
during his stay at the Institut de Recherches Subatomiques of Strasbourg.

\newpage
\renewcommand{\theequation}{A.\arabic{equation}}
\appendix
\begin{center}
\begin{Large}
{\bf Appendix }
\end{Large}
\end{center}

As mentioned in the text we gather here the expressions the ETF functionals 
take in the case of spherical symmetry. 

For the 4$^{\rm th}$ order spin independent part of the kinetic energy density, 
eq. (\ref{c2.16}), one obtains (primes denoting derivatives with respect to 
the radial variable $r$)~:
\setcounter{equation}{0}
\begin{eqnarray}
&&  \!\!\!\!\!\!\!\!
   \tau_q^{(4)}[\rho] \!=\! (3 \pi^2)^{-2/3} \frac{\rho_q^{1/3}}{4320} \left\{ 
       \frac{24}{\rho_q} \left[ \rho_q^{(4)} + \frac{4}{r} \rho_q^{'''} \right]
    - \frac{8}{\rho_q^2} \left[ 11 \rho_q^{'''} \rho_q^{'} + 7 (\rho_q^{''})^2 
    + \frac{36}{r} \rho_q^{''} \rho_q^{'} - \frac{1}{r^2} (\rho_q^{'})^2 \right]
                                                            \right. \nonumber\\
&& \;\;\;\;\;  \left. 
     + \frac{8}{3 \, \rho_q^3} \left[ 81 \rho_q^{''} (\rho_q^{'})^2 
                                          + \frac{70}{r} (\rho_q^{'})^3 \right]
     - \frac{96}{\rho_q^4} (\rho_q^{'})^4 
     - \frac{36}{f_q} \left[ f_q^{(4)} + \frac{4}{r} f_q^{'''} \right] 
     + \frac{18}{f_q^2} \left[ 4 f_q^{'''} f_q^{'}  + 
                                                    \right. \right. \nonumber\\
&& \;\;\;\;\;\left. \left. 
             + 3 (f_q^{''})^2 
             + \frac{4}{r} f_q^{''} f_q^{'} - \frac{4}{r^2} (f_q^{'})^2 \right]
     - \frac{144}{f_q^3} f_q^{''} (f_q^{'})^2  +  \frac{54}{f_q^4} (f_q^{'})^4 
     + \frac{12}{f_q\,\rho_q} \left[ 3 f_q^{'} \rho_q^{'''} 
     + 2 f_q^{''} \rho_q^{''} 
                                                    \right. \right. \nonumber\\
&& \;\;\;\;\;  \left.    \left.    
     - 2 f_q^{'''} \rho_q^{'} 
     + \frac{6}{r} f_q^{'} \rho_q^{''} - \frac{4}{r} f_q^{''} \rho_q^{'} 
     + \frac{2}{r^2} f_q^{'} \rho_q^{'} \right]
     + \frac{12}{f_q^2 \, \rho_q} \left[ f_q^{'} f_q^{''} \rho_q^{'} 
       - 2 (f_q^{'})^2 \rho_q^{''} - \frac{2}{r} (f_q^{'})^2 \rho_q^{'} \right]
                                                            \right. \nonumber\\
&&  \!\!\!\!\!\!  \left. 
     - \frac{4}{f_q \, \rho_q^2} \left[ 11 f_q^{''} (\rho_q^{'})^2 
                                      + 32 f_q^{'} \rho_q^{'} \rho_q^{''} 
                                  + \frac{26}{r} f_q^{'} (\rho_q^{'})^2 \right] 
     + \frac{30}{f_q^2 \, \rho_q^2} (f_q^{'})^2 (\rho_q^{'})^2 
     + \frac{260}{3 \, f_q \, \rho_q^3} f_q^{'} (\rho_q^{'})^3 \right\}
\label{a.1}\end{eqnarray}

When giving the spin dependent part of $\tau_q^{(4)}[\rho]$ we take advantage 
of the fact that the spin-orbit potential $\vec{W}_q$ has for Skyrme forces the 
simple form of eq.\ (\ref{c2.10}) which allows us to introduce the quantity 
\begin{equation}
  A_q = \rho + \rho_q \hbox{\hspace{1.2truecm}} \Longrightarrow  
               \hbox{\hspace{1.2truecm}} \vec{W}_q = -B_9 \vec{\nabla} A_q 
\label{a.2}\end{equation}
Using the vector identity 
$$  \mathrm{rot}(\mathrm{rot} \, \vec{a}) 
               = \vec{\nabla}(\mathrm{div} \, \vec{a}) - \Delta \, \vec{a}  $$ 
one shows that because of the form of the spin-orbit potential, eq.\ 
(\ref{c2.10}), one simply has 
$$  \vec{W} \!\cdot\! \Delta \vec{W} 
               = \vec{W} \!\cdot\! \vec{\nabla} (\mathrm{div} \vec{W}) \;   $$
which then allows us to write the spin dependent part of 
$\tau_q^{(4)_{{}_{\rm so}}}$ in the form of the local densities $\rho_n$ 
and $\rho_p$ and of their derivatives~:
\begin{eqnarray}
&&  \!\!\!\!\!\!\!\!
  \tau_q^{(4)_{{}_{\rm so}}}[\rho] \!=\! (3 \pi^{2})^{-2/3} 
        (\frac{m W_0}{\hbar^2})^2 \frac{\rho_q^{1/3}}{4f_q^2} \left\{ \! \left[ 
               \frac{3}{4} \vec{\nabla} \! A_q \!\cdot\! \vec{\nabla}^3 \! A_q
             + \frac{1}{8} \Delta(\vec{\nabla} \! A_q)^2 
             + \frac{1}{4} (\Delta A_q)^2 \right] \!
                                                            \right. \nonumber\\
&& \;\;\;\;\;  \left. 
          - \frac{1}{2f_q} \! \left[ 
               \Delta A_q (\vec{\nabla}\!f_q \!\cdot\! \vec{\nabla} \! A_q)
             + 2 \vec{\nabla} \! A_q \!\cdot\! 
                (\vec{\nabla}\! f_q \!\cdot\! \vec{\nabla}) \vec{\nabla} \! A_q
             + \vec{\nabla}\!f_q \!\cdot\! (\vec{\nabla} \! A_q \!\cdot\! 
                                              \vec{\nabla}) \vec{\nabla} \! A_q
             + \Delta f_q (\vec{\nabla} \! A_q)^2 
                                                   \right. \right. \nonumber\\
&& \;\;\;\;\;\;\;\;\;  \left. \left. 
             + \vec{\nabla} \! A_q \!\cdot\! \vec{\nabla} \! 
                              (\vec{\nabla} \! A_q \!\cdot\! \vec{\nabla}\!f_q)
             - \frac{1}{2} \vec{\nabla}\!f_q \!\cdot\! \vec{\nabla} 
                                                (\vec{\nabla} \! A_q)^2 \right] 
          + \frac{3}{4f_q^2} \left[ 
               (\vec{\nabla}\!f_q)^2 (\vec{\nabla} \! A_q)^2 
             + (\vec{\nabla}\! f_q \!\cdot\! \vec{\nabla} \! A_q)^2 
                                                   \right. \right. \nonumber\\
&& \;\;\;\;\;\;\;\;\;  \left. \left. 
             - (\frac{m W_0}{\hbar^2})^2 (\vec{\nabla} \! A_q)^4 \right] 
          + \frac{1}{6 \rho_q} \left[ 
                \vec{\nabla} \! A_q \!\cdot\! (\vec{\nabla}\!\rho_q 
                                   \!\cdot\! \vec{\nabla}) \vec{\nabla} \! A_q
             + \Delta A_q (\vec{\nabla}\!\rho_q \!\cdot\! \vec{\nabla} \! A_q) 
                                                    \right] \right. \nonumber\\
&& \;\;\;\;\;\;              
                               \!\! \left. - \frac{1}{6 f_q \rho_q} \left[ 
                    (\vec{\nabla}\!f_q \!\cdot\! \vec{\nabla}\!\rho_q) 
                                                       (\vec{\nabla} \! A_q)^2
                  + (\vec{\nabla}\!f_q \!\cdot\! \vec{\nabla} \! A_q) 
                           (\vec{\nabla}\!\rho_q \!\cdot\! \vec{\nabla} \! A_q)
                                                    \right] \right\} 
\label{a.3}\end{eqnarray}
which for spherical symmetry reads
\begin{eqnarray}
&&  \!\!\!\!\!\!\!\!\!\!
  \tau_q^{(4)_{{}_{\rm so}}}[\rho] \!=\! (3 \pi^{2})^{-2/3} \, \left( 
                 \frac{m W_0}{\hbar^2} \right)^2 \; \frac{\rho_q^{1/3}}{4f_q^2} 
                   \left\{ \frac{1}{2} \left[ 2 A_q' A_q''' + (A_q'')^2 
                      + \frac{6}{r} A_q' A_q'' - \frac{1}{r^2} (A_q')^2 \right] 
                                                           \right. \nonumber\\ 
&& \;\;\;\;\;\;\;\;\;\;\;  \left.
             - \frac{A_q'}{f_q} \left[ f_q'' A_q' 
                 + 2 f_q' A_q'' + \frac{2}{r} f_q' A_q' \right]
                    + \frac{3 (A_q')^2}{4 f_q^2} \left[ 2 (f_q')^2 
                                   - (\frac{m W_0}{\hbar^2})^2 (A_q')^2 \right]
                                                           \right. \nonumber\\
&& \;\;\;\;\;\;\;\;\;\;\;\;\;\;\;\;\;\;\;\;\;\;\;\;\;\;\;\;\;\;\;\;\;\;  \left.
              + \frac{\rho_q' A_q'}{3 f_q \rho_q} 
                                 \left[ f_q (A_q'' + \frac{1}{r} A_q')
                                           - f_q' A_q' \right] \right\} \;\; .
\label{a.4}\end{eqnarray} 
                                                                    \\[ -1.0ex]

The 4$^{\rm th}$ order spin-orbit density $\vec{J}_q^{(4)}[\rho]$, eq.\ 
(\ref{c2.28}) written in terms of the function $A_q$ defined above is given by 
\begin{eqnarray}
&&  \!\!\!\!\!\!\!\!\!\! 
   \vec{J}_q^{(4)}[\rho] = (3 \pi^{2})^{-2/3} \; \frac{m W_0}{\hbar^2} \, 
                               \frac{\rho_q^{1/3}}{8 f_q} \left\{\! 
             - 2 \vec{\nabla}^3 \! A_q
             + \frac{1}{f_q} \! \left[ \Delta f_q \, \vec{\nabla} \! A_q 
             + (\vec{\nabla} \! A_q \!\cdot\! \vec{\nabla}) \vec{\nabla} \! f_q
                                                   \right. \right.  \nonumber\\
&&  \;\;\;\;\;\;\;\;\;\;\;\;\;  \left. \left. 
             + 2 (\vec{\nabla}\! f_q \!\cdot\! \vec{\nabla}) \vec{\nabla}\! A_q
                                                                        \right] 
         - \frac{1}{f_q^2} \left[ 
               (\vec{\nabla}\! f_q)^2 \vec{\nabla}\! A_q
             + (\vec{\nabla}\!f_q \!\cdot\! \vec{\nabla}\! A_q ) 
                                                            \vec{\nabla}\! f_q 
                                                   \right. \right.  \nonumber\\
&&  \;\;\;\;\;\;\;\;\;\;\;\;\;  \left. \left.
             - 2 \, (\frac{m W_0}{\hbar^2})^2 \, (\vec{\nabla} \! A_q)^3 \right]
         - \frac{1}{3 \rho_q} \left[ 
               (\vec{\nabla}\! \rho_q \!\cdot\! \vec{\nabla}) \vec{\nabla}\! A_q
             + \Delta A_q \vec{\nabla} \! \rho_q \right]
                                                            \right. \nonumber\\
&&  \;\;\;\;\;\;\;\;\;\;  \left. 
         + \frac{1}{3 f_q \rho_q} \left[ 
                     (\vec{\nabla} \! f_q \!\cdot\! \vec{\nabla}\!\rho_q) 
                                                           \vec{\nabla} \! A_q 
             + (\vec{\nabla} \! f_q \!\cdot\! \vec{\nabla} \! A_q) 
                                                        \vec{\nabla} \! \rho_q 
                                                  \right] \right\}  
\label{c2.29}\end{eqnarray}
which in the case of a spherically symmetric system takes the form 
\begin{eqnarray}
&&  \!\!\!\!\!\!\!\!\!\! 
  \vec{J}_q^{(4)}[\rho] = (3 \pi^{2})^{-2/3} \, \frac{m W_0}{\hbar^2} \; 
                        \frac{\rho_q^{1/3}}{4f_q} \left\{
              -\!\left[ A_q''' + \frac{2}{r} A_q'' - \frac{2}{r^2} A_q' \right] 
              + \frac{1}{f_q} \left[ f_q'' A_q' + f_q' A_q'' 
                                                + \frac{1}{r} f_q' A_q' \right] 
                                                           \right. \nonumber\\ 
&&  \;\;\;\;\;\;\;\;\;\;\;\;  \left.  
              - \frac{A_q'}{f_q^2} \left[ (f_q')^2 
                                   - (\frac{m W_0}{\hbar^2})^2 (A_q')^2 \right]
              - \frac{\rho_q'}{3 f_q \rho_q} 
                                          \left[ f_q (A_q'' + \frac{1}{r} A_q')
                                       - f_q' A_q' \right] \right\} \vec{u}_r
\label{c2.30}\end{eqnarray}
with the unit vector in radial direction $\vec{u}_r$. 

The corresponding expressions for $\mathrm{div} \vec{J}_q^{(4)}[\rho]$ are 
the following~:
\begin{eqnarray*}
&&  \!\!\!\!\!\!\!\!\!\!  
  \mathrm{div} \vec{J}_q^{(4)}[\rho] \!=\! (3 \pi^{2})^{-2/3} \; 
             \frac{m W_0}{\hbar^2} \, \frac{\rho_q^{1/3}}{8 f_q} 
              \left\{ \!- 2 \Delta^2 \! A_q 
         + \frac{1}{2f_q} \left[ 2 \, \Delta f_q \, \Delta A_q 
            + \vec{\nabla}^3 \! f_q \!\!\cdot\!\! \vec{\nabla} \! A_q
                                                  \right. \right.  \nonumber\\
&& \left. \left.  \;\;\;\;\;\;  
            + 3 \Delta (\vec{\nabla}\! f_q \!\cdot\! \vec{\nabla} \! A_q)
            + 5 \vec{\nabla} \! f_q \!\!\cdot\!\! \vec{\nabla}^3 \! A_q \right] 
         - \frac{1}{f_q^2} \left[ (\vec{\nabla} \! f_q)^2 \Delta A_q 
            + 3 \Delta \! f_q (\vec{\nabla}\! f_q \!\cdot\! \vec{\nabla}\! A_q)
                                                   \right. \right.  \nonumber\\
&& \left. \left.  \;\;\;\;\;\;  
            + 5 \vec{\nabla} \! f_q \!\cdot\! (\vec{\nabla} \! A_q
                                    \!\cdot\! \vec{\nabla}) \vec{\nabla} \! f_q 
            + 5 \vec{\nabla} \! f_q \!\cdot\! 
               (\vec{\nabla} \! f_q \!\cdot\! \vec{\nabla}) \vec{\nabla} \! A_q)
            - 2 \, (\frac{m W_0}{\hbar^2})^2 
                                      \left( \Delta A_q (\vec{\nabla} \! A_q)^2 
                                           \right. \right. \right.  \nonumber\\
&& \left. \left. \left.  \;\;\;\;\;\;  
            + \vec{\nabla} \! A_q \!\cdot\! \vec{\nabla} 
                                        (\vec{\nabla} \! A_q)^2 \right) \right]
         + \frac{6}{f_q^3} \left[ (\vec{\nabla} \! f_q)^2 
                          (\vec{\nabla} \! f_q \!\cdot\! \vec{\nabla} \! A_q)
            - (\frac{m W_0}{\hbar^2})^2 \, (\vec{\nabla} \! A_q)^2
                             (\vec{\nabla} \! f_q \!\cdot\! \vec{\nabla} \! A_q)
                                                   \right] \right.  \nonumber\\
\end{eqnarray*}
\begin{eqnarray}
&& \left.  \;\;\;\;\;
         - \frac{1}{6 \rho_q} \left[ 
                  \Delta (\vec{\nabla} \! \rho_q \!\cdot\! \vec{\nabla} \! A_q)
            + 7 \vec{\nabla} \! \rho_q \!\cdot\! \vec{\nabla}^3 \! A_q
            - \vec{\nabla}^3 \! \rho_q \!\cdot\! \vec{\nabla} \! A_q
            + 2 \Delta \rho_q \, \Delta A_q
                                                   \right] \right.  \nonumber\\
&& \left.  \; 
         + \frac{1}{3 f_q \rho_q} 
                   \left[ 2 (\vec{\nabla} \! f_q \!\cdot\! \vec{\nabla} \rho_q) 
                                                                    \Delta A_q
           + \vec{\nabla} \! A_q \!\cdot\! \vec{\nabla} 
                            (\vec{\nabla} \! f_q \!\cdot\! \vec{\nabla} \rho_q) 
           + \Delta \rho_q (\vec{\nabla} \! f_q \!\cdot\! \vec{\nabla} \! A_q) +
                                                   \right. \right.  \nonumber\\
&& \left. \left.  \;\;\;\;\;\;  
             + 2 \vec{\nabla} \! \rho_q \!\cdot\! \vec{\nabla} \! 
                             (\vec{\nabla} \! f_q \!\cdot\! \vec{\nabla} \! A_q)
             + \Delta f_q (\vec{\nabla} \! \rho_q  \!\cdot\! \vec{\nabla}\! A_q)
             + 2 \vec{\nabla} \! \rho_q \!\cdot\! 
               (\vec{\nabla} \! f_q \!\cdot\! \vec{\nabla}) \vec{\nabla} \! A_q
                                                   \right] \right.  \nonumber\\
&& \left.  \; 
          - \frac{1}{3 f_q^2 \rho_q} 
                   \left[ 5 (\vec{\nabla} \! f_q \!\cdot\! \vec{\nabla} \rho_q) 
                            (\vec{\nabla} \! f_q \!\cdot\! \vec{\nabla} \! A_q)
             + (\vec{\nabla} \! f_q)^2 
                         (\vec{\nabla} \! \rho_q \!\cdot\! \vec{\nabla} \! A_q)
                                                   \right. \right.  \nonumber\\
&& \left. \left.  \;\;\;\;\;\;
             - 2 (\frac{m W_0}{\hbar^2})^2 \, (\vec{\nabla} \! A_q)^2
                (\vec{\nabla} \! \rho_q \!\cdot\! \vec{\nabla} \! A_q) \right] 
              + \frac{2}{9 \rho_q^2} 
               \left[ \vec{\nabla} \! \rho_q \!\cdot\! (\vec{\nabla} \! \rho_q
                                   \!\cdot\! \vec{\nabla}) \vec{\nabla} \! A_q
                        + (\vec{\nabla} \! \rho_q)^2 \Delta A_q
                                                   \right] \right.  \nonumber\\
&& \left.  \;
              - \frac{2}{9 f_q \rho_q^2} 
                     \left[ (\vec{\nabla} \! f_q \!\cdot\! \vec{\nabla} \rho_q) 
                         (\vec{\nabla} \! \rho_q \!\cdot\! \vec{\nabla} \! A_q)
                     + (\vec{\nabla} \! \rho_q )^2 
                            (\vec{\nabla} \! f_q \!\cdot\! \vec{\nabla} \! A_q)
                                               \right] \right\} \; . \nonumber
                                                                    \\[ -4.0ex]
\label{c2.35}\end{eqnarray}
and in the case of spherical symmetry
\begin{eqnarray}
&&  \!\!\!\!\!\!\!\!\!\! 
  \mathrm{div} \vec{J}_q^{(4)}[\rho] = (3 \pi^{2})^{-2/3} \; 
             \frac{m W_0}{\hbar^2} \, \frac{\rho_q^{1/3}}{4 f_q} 
          \left\{ -(A_q^{(4)} + \frac{4}{r} A_q''')
                            + \frac{1}{f_q} \left[ f_q''' A_q' + 2 f_q'' A_q'' 
                            + 2 f_q' A_q''' 
                                                  \right. \right.  \nonumber\\
&& \left. \left.  \;\;\;\;\;\;\;\;
                      + \frac{5}{r} f_q' A_q'' 
                       + \frac{3}{r} f_q'' A_q' - \frac{1}{r} f_q' A_q' \right]
            - \frac{1}{f_q^2} \left[ 3 (f_q')^2 A_q'' + 4 f_q' f_q'' A_q' 
                      + \frac{4}{r} (f_q')^2 A_q' 
                                                  \right. \right.  \nonumber\\
&& \left. \left.  \;\;\;\;\;\;\;\;
                           - (\frac{m W_0}{\hbar^2})^2 (A_q')^2 
                              \left( 3 A_q'' + \frac{2}{r} A_q' \right) \right]
                  + \frac{3}{f_q^3} f_q' A_q' \left[ (f_q')^2 
                                   - (\frac{m W_0}{\hbar^2})^2 (A_q')^2 \right]
                                                  \right.  \nonumber\\
&& \left.  \;\;\;
                  - \frac{1}{3 \rho_q} \left[ 2 \rho_q' A_q''' + \rho_q'' A_q'' 
                        + \frac{1}{r} \rho_q'' A_q' 
               + \frac{5}{r} \rho_q' A_q'' + \frac{1}{r^2} \rho_q' A_q' \right]
           + \frac{1}{3 f_q \rho_q} \left[ 3 f_q' \rho_q' A_q'' 
                                                  \right. \right.  \nonumber\\
&& \left. \left.  \;\;\;\;\;\;\;\;
                         + f_q' \rho_q'' A_q' + 2 f_q'' \rho_q' A_q' 
                         + \frac{4}{r} f_q' \rho_q' A_q' \right] 
           - \frac{1}{3 f_q^2 \rho_q} \rho_q' A_q' \left[ 3 (f_q')^2 
                              - (\frac{m W_0}{\hbar^2})^2 (A_q')^2 \right] 
                                                          \right.  \nonumber\\
&& \left.  \;\;\;
           +  \frac{2}{9 \rho_q^2} (\rho_q')^2 \left( A_q'' 
                                                    +  \frac{1}{r} A_q' \right)
           -  \frac{2}{9 f_q \rho_q^2} f_q' (\rho_q')^2 A_q' 
                                                            \right\} \nonumber
                                                                    \\[ -4.0ex]
\label{c2.36}\end{eqnarray}
which can also be obtained directly from eq.\ (\ref{c2.30}), remembering that 
the divergence of a vector field $\vec{a}$ that has only a radial component 
$a_r$ is given by
\[  \mathrm{div} \, \vec{a} 
                = \frac{1}{r^2} \frac{\partial}{\partial r} (r^2 a_r) \;\; . \]


\end{document}